\documentclass[11pt,a4paper]{article}   	
\usepackage{geometry}

\usepackage{graphicx}		
\usepackage{xcolor}

\usepackage{xspace}
\usepackage{subfig}

\usepackage{amsmath} 
\usepackage{ifthen}  
\usepackage{amssymb}
\usepackage{amsfonts}
\usepackage{upgreek} 

\usepackage{listings}

\usepackage[T1]{fontenc}
\usepackage{bm}

\usepackage{lineno}
\usepackage{calc}

\usepackage{hyperref}
\usepackage{cancel}
\newboolean{uprightparticles}
\setboolean{uprightparticles}{false}

\newcommand{\tev}{\ensuremath{\mathrm{\,Te\kern -0.1em V}}\xspace}
\newcommand{\gev}{\ensuremath{\mathrm{\,Ge\kern -0.1em V}}\xspace}
\newcommand{\mev}{\ensuremath{\mathrm{\,Me\kern -0.1em V}}\xspace}
\newcommand{\kev}{\ensuremath{\mathrm{\,ke\kern -0.1em V}}\xspace}
\newcommand{\ev}{\ensuremath{\mathrm{\,e\kern -0.1em V}}\xspace}
\newcommand{\gevc}{\ensuremath{{\mathrm{\,Ge\kern -0.1em V\!/}c}}\xspace}
\newcommand{\mevc}{\ensuremath{{\mathrm{\,Me\kern -0.1em V\!/}c}}\xspace}
\newcommand{\gevcc}{\ensuremath{{\mathrm{\,Ge\kern -0.1em V\!/}c^2}}\xspace}
\newcommand{\gevgevcccc}{\ensuremath{{\mathrm{\,Ge\kern -0.1em V^2\!/}c^4}}\xspace}
\newcommand{\mevcc}{\ensuremath{{\mathrm{\,Me\kern -0.1em V\!/}c^2}}\xspace}

\newcommand{\eg}{\mbox{\itshape e.g.}\xspace}
\newcommand{\ie}{\mbox{\itshape i.e.}}

\newcommand{\Bd}{\ensuremath{B^0}}
\newcommand{\Kstarz}{\ensuremath{K^{*0}}}
\newcommand{\mumu}{\ensuremath{\mu^+\mu^-}}
\newcommand{\thetal}{\ensuremath{\theta_l}}
\newcommand{\thetak}{\ensuremath{\theta_K}}
\newcommand{\ctl}{\ensuremath{\cos\thetal}}
\newcommand{\ctk}{\ensuremath{\cos\thetak}}
\newcommand{\fsig}{\ensuremath{f_{\rm sig}}}

\newboolean{articletitles}
\setboolean{articletitles}{true}

\usepackage{cite}
\usepackage{mciteplus}

\usepackage{enumitem}

\usepackage{array,booktabs,ragged2e}
\newcolumntype{R}[1]{>{\RaggedLeft\arraybackslash}p{#1}}

\usepackage{etoolbox}

\newboolean{inbibliography}

\newcommand*\linenomathpatchAMS[1]{%
  \expandafter\pretocmd\csname #1\endcsname {\linenomathAMS}{}{}%
  \expandafter\pretocmd\csname #1*\endcsname{\linenomathAMS}{}{}%
  \expandafter\apptocmd\csname end#1\endcsname {\endlinenomath}{}{}%
  \expandafter\apptocmd\csname end#1*\endcsname{\endlinenomath}{}{}%
}

\expandafter\ifx\linenomath\linenomathWithnumbers
  \let\linenomathAMS\linenomathWithnumbers
  \patchcmd\linenomathAMS{\advance\postdisplaypenalty\linenopenalty}{}{}{}
\else
  \let\linenomathAMS\linenomathNonumbers
\fi

\linenomathpatchAMS{gather}
\linenomathpatchAMS{multline}
\linenomathpatchAMS{align}
\linenomathpatchAMS{alignat}
\linenomathpatchAMS{flalign}

\newcommand{\morefit}{\mbox{MoreFit}}

\newcommand{\rootsystem}{\texttt{ROOT}}
\newcommand{\mycomment}[1]{}
  
\definecolor{verylightgray}{rgb}{0.95,0.95,0.95}
  
\begin{document}
\pagenumbering{roman}
\thispagestyle{empty}

\vspace*{1.0cm}
\begin{center}
{\huge\bfseries \boldmath
\morefit\\[0.2cm] \large A More Optimised, Rapid and Efficient Fit}\\[1.0 cm]
{\Large
Christoph~Langenbruch$^{a}$, 
}\\[0.4 cm] 
{\small
$^a$ Physikalisches Institut, Heidelberg University, Im Neuenheimer Feld 226, 69120 Heidelberg, Germany
} \\[0.5 cm]
\small
E-Mail:
\texttt{\href{mailto:christoph.langenbruch@cern.ch}{christoph.langenbruch@cern.ch}}.
\vspace{\fill}

\vspace*{2.0cm}

{\normalsize\textsc{Keywords:} Parallel computation, Maximum likelihood, Heterogeneous architectures, GPGPU, OpenCL, LLVM, Clang, Just-in-time compilation, Computation graphs, Automatic optimisation, Symbolic differentiation}

\vspace*{1.0cm}

\begin{abstract} 
  \noindent 
  Parameter estimation via unbinned maximum likelihood fits is 
  a central technique in particle physics. 
  This article introduces \morefit, which aims to provide a more optimised, rapid and efficient fitting solution for unbinned maximum likelihood fits. 
  \morefit\ is developed with a focus on parallelism and relies on computation graphs that are 
  compiled just-in-time. 
  Several novel automatic optimisation techniques are employed on the computation graphs that significantly increase performance compared to conventional approaches. 
  \morefit\ can make efficient use of a wide range of heterogeneous platforms through its compute backends that rely on open standards. 
  It provides an OpenCL backend for execution on GPUs of all major vendors,
  and a backend based on LLVM and Clang for single- or multithreaded execution on CPUs, which in addition allows for SIMD vectorisation.
  \morefit\ is benchmarked against several other fitting frameworks and shows very promising performance,
  illustrating the power of the approach. 
\end{abstract} 

\vspace*{1.75cm}

\begin{center}
  Published in Eur.~Phys.~J.~C \textbf{86} (2026) 105
\end{center}

\vspace{\fill}

{\footnotesize 
\centerline{\copyright~C.\ Langenbruch, licence \href{http://creativecommons.org/licenses/by/4.0/}{CC-BY-4.0}.}}
\end{center}

\clearpage

\setcounter{page}{1}
\pagenumbering{arabic}

\section{Introduction} 
\enlargethispage{1em}
\label{sec:introduction}
The estimation of parameters and 
their confidence intervals from data is a central task in the physical sciences.
In particle physics, unbinned maximum likelihood fits are an essential tool for parameter inference, 
as, in the asymptotic limit, the maximum likelihood estimator is normally distributed around the true parameter value and its variance is equal to the minimum variance bound~\cite{James:2006zz}. 
Furthermore, unbinned fits do not lose information due to binning. 
However, the unprecedented amount of data taken in modern particle physics experiments pose computational challenges. 
It is not unusual for data samples to contain~${\cal O}(10^6)$ events from which more than $100$ parameters need to be determined.
Even the analysis of smaller data samples can be time- and energy-consuming if techniques for coverage correction need to be employed~\cite{Neyman:1937uhy,Feldman:1997qc}. 
These methods typically rely on large numbers of pseudo-experiments being performed.
For each parameter which requires coverage correction, often more than ${\cal O}(10^5)$ pseudo-experiments are needed. 

To address these computational challenges, 
an obvious strategy is to exploit 
the ``embarrassingly parallel'' nature of the likelihood computation:  
Since the determination of the logarithmic probability density function for each event is independent,
the problem can be easily parallelised over each event. 
Several existing fitting frameworks implement parallel computation for unbinned likelihoods, 
among them RooFit~\cite{Verkerke:2003ir}, zfit~\cite{Eschle:2019jmu}, GooFit~\cite{Andreassen_2014,Schreiner:2017csm}, and the TensorFlowAnalysis package~\cite{TensorFlowAnalysis}. 
These software packages typically rely on CUDA~\cite{Nickolls:2008gqs}, either directly or indirectly via the TensorFlow library~\cite{tensorflow2015-whitepaper}, to allow execution on NVIDIA GPUs. 

This article introduces \morefit, 
a new fitting framework for unbinned maximum likelihood fits developed 
with a focus on parallelism and automatic optimisation. 
\morefit\ is based on automatically optimised computation graphs that are compiled just-in-time. 
The determination of the likelihood can be performed on heterogeneous platforms:
The default compute backend uses the open and vendor-independent OpenCL standard which is supported by GPUs of all major vendors~\cite{OpenCL}.
For computation on CPUs, an additional backend based on LLVM~\cite{LLVM:CGO04} and Clang~\cite{clang} is available, which allows for vectorised single- and multithreaded execution. 
\morefit\ is a \texttt{C++} only library which aims to be lightweight with minimal dependencies. 
As such it does not rely on TensorFlow or \rootsystem~\cite{Brun:1997pa}, though compilation with the \rootsystem\ libraries is possible to enable \rootsystem-related functionality. 

This article is structured as follows: 
Section~\ref{sec:maximumlikelihood} briefly introduces the problem of parameter estimation using unbinned maximum likelihood fits and Sec.~\ref{sec:computationgraphs} details how \morefit\ approaches this problem using computation graphs that allow for automatic optimisation.
Technical details on the compute backends are given in Sec.~\ref{sec:backends}. 
Sections~\ref{sec:optimisations} and~\ref{sec:generation} discuss parameter fits and event generation, 
as well as 
the automatic optimisations that are performed by \morefit\ to be as efficient as possible in these areas. 
\morefit\ provides all of the discussed optimisations automatically, for both built-in as well as user-supplied PDFs. 
The resulting performance of \morefit\ is studied in Sec.~\ref{sec:benchmarking}, where it is benchmarked against RooFit~\cite{Verkerke:2003ir} and zfit~\cite{Eschle:2019jmu} using two simple examples. 
Section~\ref{sec:summary} gives an outlook on future developments of \morefit\ and provides a summary. 

\section{Maximum likelihood fits and the \morefit\ strategy}
\subsection{Maximisation of unbinned likelihood}
\label{sec:maximumlikelihood}
Unbinned maximum likelihood fits are based on the minimisation of the negative logarithmic likelihood,
\begin{align}
  -\ln {\cal L}(\bm{x}_1,\ldots,\bm{x}_N|\bm{\lambda}) &= -\sum_{i=1}^{N} \ln{\cal P}(\bm{x}_i|\bm{\lambda}),
\end{align}
where ${\cal P}$ denotes the probability density function (PDF),
evaluated for the $N$ data points $\bm{x}_{i=1,\ldots,N}$, and the parameter set $\bm{\lambda}$. 
Since $\ln{\cal P}(\bm{x}_i|\bm{\lambda})$ is independent for all $i$,
the calculation of the (often computationally expensive) logarithmic PDF 
can be parallelised over the $N$ events.
Minimisation of the negative logarithmic likelihood requires its repeated evaluation at many parameter points.
In addition, gradient-based minimisers can often be more efficient than gradient-free methods. 
Therefore, a fast and accurate determination of both the logarithmic likelihood as well as its partial derivatives is necessary. 
For the determination of parameter uncertainties, the second derivatives of the logarithmic likelihood are needed in addition. 
Being able to use analytic derivatives can be very beneficial in this context, as they tend to be more stable than numerical approximations and can be faster. 
\morefit\ provides the analytic derivatives via symbolic differentiation of a computation graph, as detailed below in Sec.~\ref{sec:computationgraphs}. 

In high energy physics, the most popular software used for minimisation of the negative logarithmic likelihood is the \texttt{Minuit} software package~\cite{James:1975dr,James:1994vla}.
Its minimiser \texttt{Migrad} uses by default the Davidon-Fletcher-Powell variable-metric algorithm~\cite{FletcherPowell} or alternatively the Broyden–Fletcher–Goldfarb–Shanno algorithm~\cite{Broyden,Fletcher,Goldfarb,Shanno}. 
\morefit\ uses the \texttt{Minuit2} standalone minimiser which can be used without compilation with the \rootsystem\ libraries~\cite{Brun:1997pa}. 
When the \rootsystem\ libraries are available, the legacy \texttt{TMinuit} minimiser can be used alternatively.
While in \morefit\ the minimisation algorithm is always executed on the host machine,
the computationally expensive determination of the logarithmic likelihood can be performed on an accelerator, such as a GPU. 
This involves data transfer from the host to the accelerator which can incur some overhead.
Care should therefore be taken to minimise these transfers as much as possible. 
The minimisation process repeatedly evaluates the logarithmic likelihood and, if requested, the analytic gradient for the same data sample at different parameter values.
This opens up several opportunities for optimisations, as will be discussed in Sec.~\ref{sec:optimisations}. 

\subsection{Computation graphs}
\label{sec:computationgraphs}
\begin{figure}
  \centering
  \includegraphics[width=0.3\linewidth]{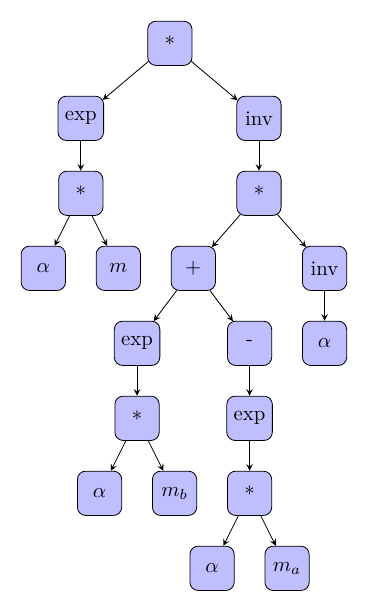}
  \caption{Computation graph for a simple normalised exponential PDF. Note that the graph is slightly simplified for illustration purposes, as \morefit\ by default also handles the special case $\alpha=0$.\label{fig:exponential}}
\end{figure}

PDFs in \morefit\ are implemented using computation graphs,
from which compute kernels are written automatically that are compiled just-in-time. 
Computation graphs store a calculation in a simple tree structure, with nodes consisting of basic operations, functions, variables and constants. 
Each PDF, either built-in or user-supplied, needs to implement at least the unnormalised probability density and the corresponding analytic normalisation,
from which the computation graph for the logarithmic normalised probability density can be obtained. 
As an example, Fig.~\ref{fig:exponential} gives the computation graph for an exponential function ${\rm exp}(\alpha m)$ normalised over the range $m\in [m_a, m_b]$. 
Computation graphs can provide solutions to several of the challenges for fitting frameworks.
In particular, they allow for easy symbolic differentiation via the chain rule and can be automatically analysed and optimised, depending on the use case. 

As an example for symbolic differentiation, the derivative of the graph in Fig.~\ref{fig:exponential} with respect to the slope parameter $\alpha$ is shown in Fig.~\ref{fig:derivativeexponential} in the Appendix. 
Note that no optimisations have been applied by \morefit\ for these graphs. 
If requested, \morefit\ uses symbolic differentiation to calculate the gradient of the logarithmic likelihood and the Hessian matrix,
\ie\ the matrix of the second derivatives of the logarithmic likelihood. 
In addition, also the Fisher information matrix can be calculated analytically, which is required for weighted fits~\cite{Langenbruch:2019nwe}. 

Concerning automatic optimisations, 
\morefit\ initially performs some trivial simplifications, including the merging and evaluation of all nodes that are constant a priori. 
However, more advanced optimisations are possible that can result in considerable performance gains both
during the fitting process as described in Sec.~\ref{sec:optimisations} and 
in the generation of pseudoexperiments as discussed in Sec.~\ref{sec:generation}. 

\subsection{Computation backends}
\label{sec:backends}
While the minimisation algorithm used by \morefit\ is always executed on the host, 
the calculation of the logarithmic likelihood, and, if requested, its analytic gradient as well as the Hessian at the parameter point $\bm{\lambda}$ will be performed on an accelerator.
The accelerator, which can be a GPU or alternatively one or multiple CPUs on the host, will perform this calculation parallelising over the events. 
It is the responsibility of the compute backends to write the kernels for the specific accelerator from the computation graph,
after the computation graph has been optimised as will be detailed in the following. 
Furthermore, the backend has to provide the necessary interface for the data to be transferred to and from the accelerator.
The compute kernels are compiled just-in-time before execution, which incurs some computational overhead at run-time.
This overhead will only be significant for small data sets and reduces for large samples. 
However, for small data sets the only computationally expensive task is typically the repeated generation and fit of pseudoexperiments for fit validation or coverage correction. 
In these cases the kernels only have to be compiled once, and then can be run repeatedly (typically ${\cal O}(1000)$ times),
such that the overhead from the kernel compilation significantly reduces. 
The default \morefit\ backend for performing computations on GPUs is based on OpenCL,
the default compute backend on CPUs is based on LLVM. 
Details are given in Secs.~\ref{sec:opencl} and~\ref{sec:llvm} below. 

\subsubsection{OpenCL backend}
\label{sec:opencl}
\morefit\ relies on OpenCL~\cite{OpenCL}, a vendor-independent open standard, to be able to use GPUs of all major vendors. 
The backend writes small OpenCL~C kernels 
for likelihood evaluation, determination of the gradient and Hessian, as well as for the event generation. 
The likelihood evaluation takes as input the parameters~$\bm{\lambda}$ at which the logarithmic likelihood has to be evaluated and the data in the event variables. 
The data are stored as a structure of arrays in the event variables and padded to a multiple of the chosen work group size\footnote{OpenCL Kernels are executed in parallel as work items, grouped in work groups. The global work size, \ie\ the total number of work items, must be a multiple of the work group size. The work group size can be chosen a priori, but may be limited by hardware constraints.}. 
For each event the output of the likelihood kernel is the normalised logarithmic PDF evaluated for the specific event. 
Listing~\ref{list:openclkernel} shows as an example a kernel for the evaluation of a simple exponential PDF.
Note that all optimisations beyond trivial simplifications for this kernel were switched off for illustration purposes, 
Secs.~\ref{sec:parameteroptimisation} and~\ref{sec:pereventoptimisation} will detail how the kernel can be automatically optimised to be more efficient.
After the calculation of all work items is completed, the summation over all individual logarithmic probability densities needs to be performed to obtain the total logarithmic likelihood.
This can either be done on the host, in which case the results are transferred to the host where a Kahan summation is performed~\cite{KahanSummation}.
Alternatively, the reduction can be performed on the accelerator.
To this end, multiple reduction kernels are submitted in parallel that perform Kahan summation of the logarithmic probabilities with a configurable reduction factor.
This process is repeated until the number of terms to add is small enough that only one kernel needs to be run which then returns the total logarithmic likelihood.
To reduce data transfers between host and accelerator, the Kahan summation on the accelerator is the default option in \morefit. 

\begin{lstlisting}[float,caption={Example for an OpenCL kernel to calculate the logarithm of a simple exponential PDF written by \morefit. \morefit\ automatically performs the trivial simplification $\ln(\exp(\alpha m))=\alpha m$, but for illustration purposes no further optimisations are applied here.},captionpos=b,label=list:openclkernel]
__kernel void lh_kernel(__const int nevents, __const int nevents_padded, __global const double* data, __global double* output, __global const double* parameters)
{
int idx = get_global_id(0); //event index
const double m = data[idx]; //mass of event
const double alpha = parameters[0];
output[idx] = ((alpha*m)+-(log(((alpha==0.0) ? 2.000000000000000e+00 : ((exp((7.000000000000000e+00*alpha))+-(exp((5.000000000000000e+00*alpha))))*1.0/(alpha)))))); //alpha*m-log(norm)
}
\end{lstlisting}

\subsubsection{LLVM backend}
\label{sec:llvm}
The default compute backend for performing computations on CPUs uses C kernels compiled just-in-time by LLVM and Clang. 
Several options can be set that control the optimisation level of the compilation. 
The event loop is structured such that it allows to perform the calculations on multiple events at the same time, \ie\ Single Instruction Multiple Data (SIMD), via auto-vectorisation. 
The vectorisation width can be set depending on the CPU architecture.
Typical modern CPUs support a vectorisation width of four, \ie\ four operations can be performed simultaneously in double precision.
Auto-vectorisation is enabled in \morefit\ by default. 
For the benchmarks performed in Secs.~\ref{sec:massfit} and \ref{sec:angularfit}, auto-vectorisation results 
in speed-ups corresponding to factors of up to $2.4$ and $1.9$, respectively. 
In addition to vectorisation, the LLVM backend allows to split the events in blocks to allow for multithreaded execution. 
Repeated creation and destruction of threads would induce significant overhead.
Therefore, a simple thread pooling strategy is used, 
where during the minimisation the threads are kept alive and reused for each iteration of the parameters~$\bm{\lambda}$. 
For the LLVM backend, Kahan summation~\cite{KahanSummation} (vectorised if requested) is used throughout.

\subsection{Parameter fits and automatic optimisations}
\label{sec:optimisations}
\subsubsection{Optimisation of parameter-dependent terms}
\label{sec:parameteroptimisation}
\begin{figure}
\centering
\includegraphics[width=0.95\linewidth]{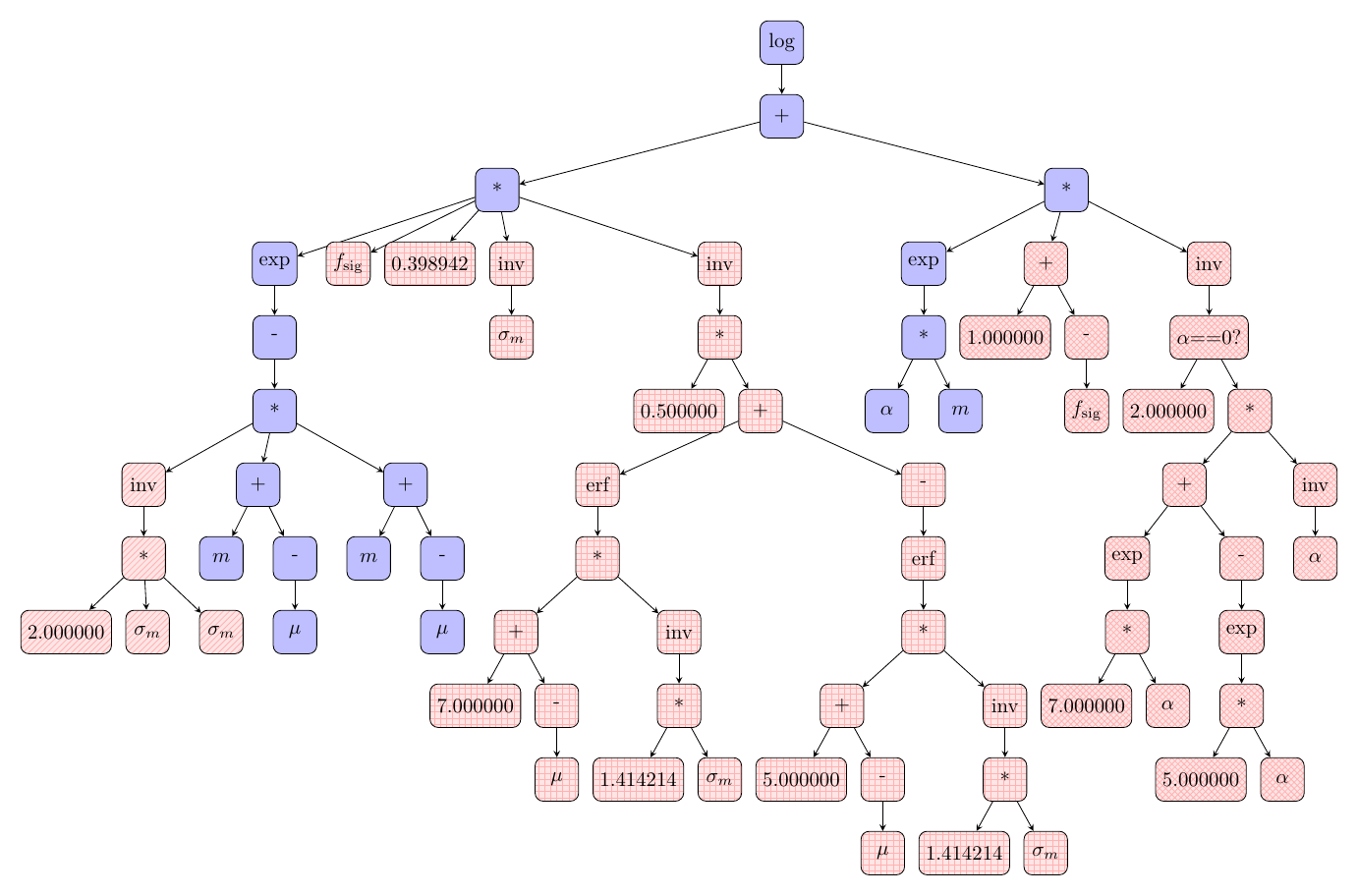}
\caption{Unoptimised computation graph for the logarithm of the model given by Eq.~\ref{eq:massmodel}. 
  Terms depending only on parameters (and not on the event) can be calculated once per parameter set on the host and reused. 
  These terms are automatically identified by \morefit\ and are highlighted above in light red with different hatching styles.
  For unbinned maximum likelihood fits they include the normalisation integrals. 
  \label{fig:probgraph}}
\end{figure}

Figure~\ref{fig:probgraph} shows the computation graph for 
a simple model, consisting of a mixture of a Gaussian and an exponential component. 
The total PDF, normalised over the range $m\in[m_a,m_b]$, is given by
\begin{align}
  {\cal P}(m;f_{\rm sig},\mu,\sigma_m,\alpha) &= f_{\rm sig}\times \frac{1}{\sqrt{2\pi}\sigma_m} e^{-\frac{(m-\mu)^2}{2\sigma_m^2}}/{\cal N}_1 + (1-f_{\rm sig}) e^{\alpha m}/{\cal N}_2~~~{\text{with}}\label{eq:massmodel}\\
  {\cal N}_1(\mu,\sigma_m) &= \frac{1}{2}\biggl({\rm erf}\bigl(\tfrac{\mu-m_a}{\sqrt{2}\sigma_m}\bigr)-{\rm erf}\bigl(\tfrac{\mu-m_b}{\sqrt{2}\sigma_m}\bigr)\biggr)~~~{\text{and}}\nonumber\\
  {\cal N}_2(\alpha) &=
  \left\{\begin{array}{ll}
  m_b-m_a & ~\text{for}~\alpha=0\\
  \left(e^{\alpha m_b}-e^{\alpha m_a}\right)/\alpha\nonumber & ~\text{otherwise},
\end{array}\right.
\end{align}
  where $f_{\rm sig}$ denotes the signal fraction, $\mu$ ($\sigma_m$) the Gaussian mean (width), and $\alpha$ the exponential slope. 
In this example, the range was chosen to be $m\in [5,7]\gevcc$. 
When analysing computation graphs such as Fig.~\ref{fig:probgraph} it becomes apparent that there are terms that only depend on the parameters and not the specific event.
In Fig.~\ref{fig:probgraph} such terms are coloured in light red. 
Examples of such terms include the normalisations of the PDFs which by definition do not depend on the event. 
A straightforward optimisation consists in calculating these terms only once on the host for each parameter update and buffer the result to be used for each event. 
For the normalisation, which often can be computationally costly, this is a well-known technique for optimisation and used for example in the RooFit~\cite{Verkerke:2003ir} fitting framework. 
In \morefit\ the computation graphs are automatically analysed and searched for nodes that do not depend on event variables, if multiple such terms appear in sums or products these terms are combined. 
If such terms exceed a specified computational cost they are automatically optimised and the buffered result is then used in the compute kernel. 
For the example in Fig.~\ref{fig:probgraph}, three terms are identified that can be calculated on the host:
\begin{align}
  b_0 &= \frac{1}{2\sigma_m^2}\\
  b_1 &= f_{\rm sig} \frac{1}{\sqrt{2\pi}\sigma_m} \bigg/\biggl[\frac{1}{2}\biggl({\rm erf}\bigl(\tfrac{\mu-m_a}{\sqrt{2}\sigma_m}\bigr)-{\rm erf}\bigl(\tfrac{\mu-m_b}{\sqrt{2}\sigma_m}\bigr)\biggr)\biggr]\nonumber\\
  b_2 &= \left(1-f_{\rm sig}\right) \big/
  \left\{\begin{array}{ll}
  m_b-m_a & ~\text{for}~\alpha=0\\
  \left(e^{\alpha m_b}-e^{\alpha m_a}\right)/\alpha\nonumber & ~\text{otherwise},
\end{array}\right.\nonumber
\end{align}
These terms correspond to the denominator in the Gaussian PDF and the two normalisations, which are combined with other parameter-dependent factors. 
As illustrated in Listings~\ref{list:unoptkernel} and~\ref{list:optkernel}, which show the unoptimised and the resulting optimised OpenCL kernel,
the buffered values are transmitted as additional parameters to the (overall much smaller) optimised kernel. 
\begin{lstlisting}[float,caption={OpenCL kernel to calculate the logarithm of the PDF given by Eq.~\ref{eq:massmodel}, written by \morefit\ without optimisations.},captionpos=b,label=list:unoptkernel]
__kernel void lh_kernel(__const int nevents, __const int nevents_padded, __global const double* data, __global double* output, __global const double* parameters)
{
int idx = get_global_id(0);
const double m = data[idx];
const double mu = parameters[0];
const double sigma = parameters[1];
const double fsig = parameters[2];
const double alpha = parameters[3];
output[idx] = log(((3.989422804014327e-01*fsig*1.0/((5.000000000000000e-01*(erf(((7.000000000000000e+00+-(mu))*1.0/((1.414213562373095e+00*sigma))))+-(erf(((5.000000000000000e+00+-(mu))*1.0/((1.414213562373095e+00*sigma))))))))*exp(-((1.0/((2.000000000000000e+00*sigma*sigma))*(m+-(mu))*(m+-(mu)))))*1.0/(sigma))+((1.000000000000000e+00+-(fsig))*exp((alpha*m))*1.0/(((alpha==0.0) ?2.000000000000000e+00 : ((exp((7.000000000000000e+00*alpha))+-(exp((5.000000000000000e+00*alpha))))*1.0/(alpha)))))));
}
\end{lstlisting}
\begin{lstlisting}[float,caption={OpenCL kernel to calculate the logarithm of the PDF given by Eq.~\ref{eq:massmodel}, written by \morefit\ when automatically optimising parameter-dependent terms.},captionpos=b,label=list:optkernel]
  __kernel void lh_kernel(__const int nevents, __const int nevents_padded, __global const double* data, __global double* output, __global const double* parameters)
{
int idx = get_global_id(0);
const double m = data[idx];
const double mu = parameters[0];
const double sigma = parameters[1];
const double fsig = parameters[2];
const double alpha = parameters[3];
const double morefit_parambuffer_0 = parameters[4];
const double morefit_parambuffer_1 = parameters[5];
const double morefit_parambuffer_2 = parameters[6];
output[idx] = log(((exp(-(((m+-(mu))*(m+-(mu))*morefit_parambuffer_0)))*morefit_parambuffer_1)+(exp((m*alpha))*morefit_parambuffer_2)));
}
\end{lstlisting}

The reader may notice that the expression $m-\mu$ appears twice in Fig.~\ref{fig:probgraph}, 
and therefore should only be calculated once and reused. 
Optimisations of this kind are known as Common Subexpression Elimination (CSE), and are a common technique used by compilers. 
Since the expressions are used in the same block in the kernel the compiler should be able to perform this optimisation automatically. 
The intermediate representation produced by LLVM confirms that this is indeed the case here.

\subsubsection{Optimisation of event-dependent terms}
\label{sec:pereventoptimisation}
Besides terms that only depend on parameters but not on event variables it can also be beneficial to study terms that only depend on event variables. 
To better appreciate why this can result in optimisations it is useful to consider the example of a typical angular fit in flavour physics, such as the angular analysis of the rare decay $\Bd\to\Kstarz\mumu$~\cite{LHCb:2015svh,LHCb:2020lmf}.
The PDF describing the differential decay rate of the signal depends on the three decay angles $\cos\thetal$, $\cos\thetak$ and $\phi$, and is given by~\cite{Altmannshofer:2008dz}
\begin{align}
\frac{{\rm d}^3\Gamma}{{\rm dcos}\theta_l\,{\rm dcos}\theta_K\,{\rm d}\phi} =& \sum_{i=1}^{11} {\color{blue}S_i} {\color{red}f_i(\ctl,\ctk,\phi)}\label{eq:angulardecayrate}\\
=& \frac{9}{32\pi}
\bigl[ (1-{\color{blue}F_{\rm L}}){\color{red}\tfrac{3}{4} \sin^2\thetak} + {\color{blue}F_{\rm L}}{\color{red}\cos^2\thetak} + (1-{\color{blue}F_{\rm L}}) {\color{red}\tfrac{1}{4} \sin^2\thetak\cos 2\thetal}\nonumber\\
& \hphantom{\frac{9}{32\pi}} - {\color{blue}F_{\rm L}} {\color{red}\cos^2\thetak\cos 2\thetal} + {\color{blue}S_3}{\color{red}\sin^2\thetak \sin^2\thetal \cos 2\phi}\nonumber\\
& \hphantom{\frac{9}{32\pi}} + {\color{blue}S_4} {\color{red}\sin 2\thetak \sin 2\thetal \cos\phi} + {\color{blue}S_5}{\color{red}\sin 2\thetak \sin \thetal \cos \phi}\nonumber\\
& \hphantom{\frac{9}{32\pi}} + {\color{blue}A_{\rm FB}} {\color{red}\tfrac{4}{3}\sin^2\thetak \cos\thetal} + {\color{blue}S_7} {\color{red}\sin 2\thetak \sin\thetal \sin\phi}\nonumber\\
& \hphantom{\frac{9}{32\pi}} + {\color{blue}S_8} {\color{red}\sin 2\thetak \sin 2\thetal \sin\phi} + {\color{blue}S_9}{\color{red}\sin^2\thetak \sin^2\thetal \sin 2\phi}
\bigr].\nonumber
\end{align}
The objective is to fit the angular coefficients given in blue that appear in front of the angular terms $f_i(\ctl,\ctk,\phi)$ given in red.
The angular terms notably only depend on the event and not on the parameters. 
During the minimisation process the logarithmic likelihood will have to be evaluated for the same data at multiple different parameter points. 
It can thus be beneficial to evaluate the angular terms $f_i(\ctl,\ctk,\phi)$ for each event only once, in a precomputation step. 
The precomputation step will increase the dimensionality of the data from $\bm{x}_i=\{\cos\theta_{l,i}, \cos\theta_{K,i}, \phi_i\}$ to $\bm{x}_i^\prime=\{\cos\theta_{l,i}, \cos\theta_{K,i}, \phi_i,\allowbreak f_{1,i}, \ldots, f_{11,i}\}$.
The minimisation process can afterwards simply use the buffered values for $f_{1,i},\ldots,f_{11,i}$ instead of recomputing them at each parameter step.
While this approach increases memory pressure, it can avoid the repeated, potentially computationally expensive, evaluation of the same expressions. 
Depending on the problem this can be a very powerful optimisation technique. 
For the angular fit discussed here, the relative performance improvement is discussed in Sec.~\ref{sec:angularfit}. 

\morefit\ is able to perform this optimisation completely automatically. 
First, the computation graph is traversed, searching for terms depending only on the event that exhibit a computational cost above a user-specified threshold. 
Then, a specific kernel is written and launched to calculate the new data set of higher dimensionality. 
Finally, a simplified kernel is written which uses the buffered per-event quantities. 
For the example above, these kernels are given for illustration in Listings~\ref{list:precomputationkstar} and~\ref{list:pereventkstar} in the Appendix.

\subsection{Fast generation of pseudoexperiments}
\label{sec:generation}
The generation of pseudodata is essential for fit validation to ensure that the fit is unbiased and the provided uncertainties have the correct coverage. 
In addition, pseudodata is needed for coverage correction methods that perform a Neyman construction~\cite{Neyman:1937uhy},
such as the Feldman-Cousins method~\cite{Feldman:1997qc}. 
It is therefore important for any fitting framework to not only efficiently fit data, but also to be able to rapidly generate pseudodata according to a given PDF. 

\begin{figure}
  \centering
  \subfloat[mass generation graph\label{fig:massgeneration}]{\includegraphics[height=0.4\linewidth]{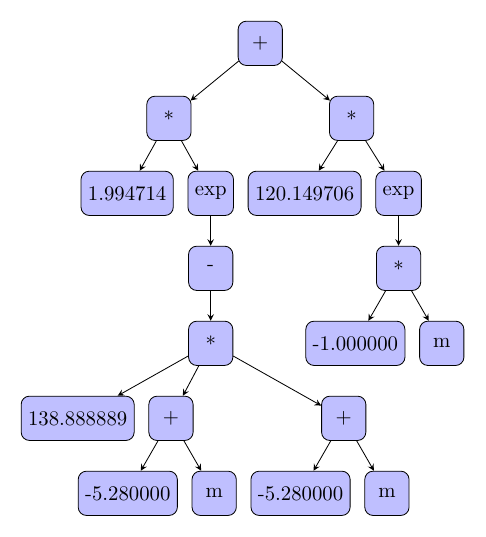}}
  \hspace{0.02\linewidth}
  \subfloat[angular generation graph\label{fig:angulargeneration}]{\includegraphics[height=0.36\linewidth]{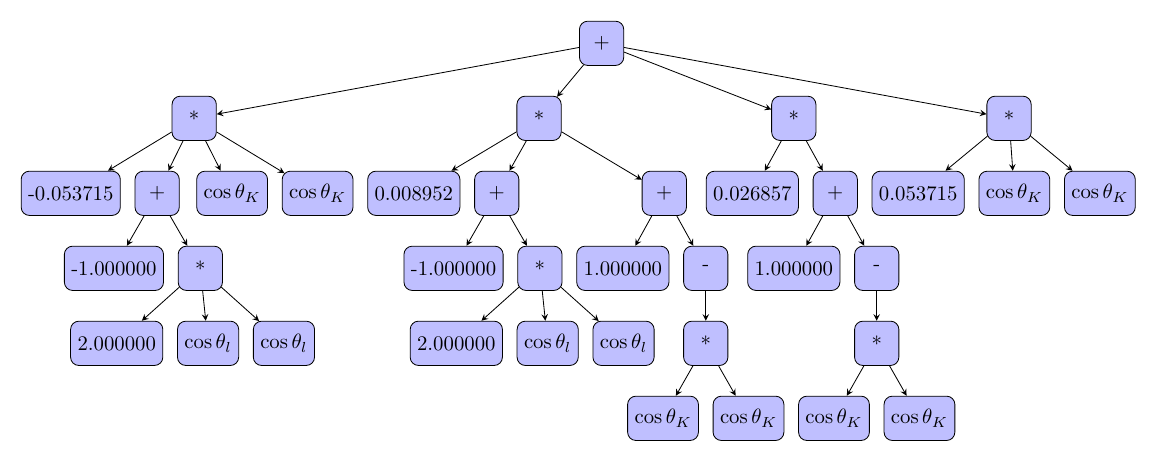}}
  \caption{(a) Computation graph used in the generation for 
    the model corresponding to Eq.~\ref{eq:massmodel}. 
  All terms depending on parameters are constants in the generation and can thus be evaluated in advance, resulting in a much reduced computation graph size compared to Fig.~\ref{fig:probgraph} and a smaller and faster kernel.
  (b) Computation graph for the generation of the decay angles discussed in Sec.~\ref{sec:angularfit}. In the generation, $F_{\rm L}$ is set to $0.6$ and all other angular observables are set to zero. As a result, only the first four terms in Eq.~\ref{eq:angulardecayrate} remain and the generation is flat in the angle $\phi$. 
  \label{fig:gengraph}}
\end{figure}

\morefit\ aggressively optimises the generation of pseudodata by exploiting the fact that all parameters in the generation are fixed. 
Terms in the computation graph that only depend on parameters and not the event variables that need to be generated are therefore constants that are evaluated in advance. 
As an example, Fig.~\ref{fig:massgeneration} shows the computation graph for the model given by Eq.~\ref{eq:massmodel}. 
Compared to the full PDF in Fig.~\ref{fig:probgraph} the number of nodes significantly reduces, resulting in a significant speedup in the generation. 
Figure~\ref{fig:angulargeneration} shows the graph for the generation of the decay angles according to the PDF given by Eq.~\ref{eq:angulardecayrate}.
In this case, the parameter $F_{\rm L}$ is set to $0.6$ and all other angular observables are set to zero in the generation.
As a consequence, only the first four terms in Eq.~\ref{eq:angulardecayrate} remain
and the other angular terms are removed by the automatic optimisation, resulting in a significant simplification. 

The generation itself is implemented in \morefit\ via the accept-reject approach. 
While there are more optimised approaches for generation possible for simple PDFs, 
\eg\ by using the inverse of the cumulative distribution function, this is left as a potential further optimisation.
The generation of pseudoexperiments relies on pseudo random number generators. 
\morefit\ allows generation of pseudo random numbers through several algorithms,
currently implemented generators include the Mersenne-Twister~\cite{MersenneTwister}, \texttt{Xoshiro128++/256++}~\cite{Xoshiro}, and \texttt{PCG64DXSM}~\cite{oneill:pcg2014}. 
Two options are available for the generation of events via the accept-reject method. 

The first option consists of generating events uniformly in the event variables on the host and transferring them to the accelerator where their probability is calculated.
The results are then transferred back to the host where the events are either accepted or rejected, depending on their probability.
This is repeated until the required number of events is generated. 
The repeated data transfers between host and accelerator do incur some overhead which can slow the generation process down. 

As an alternative, \morefit\ therefore allows the full generation process, including the random number generation and the accept/reject step, to take place on the accelerator.
This is only possible for pseudo random number generators with a comparably small internal state as there are limitations on the numbers of available registers on GPUs. 
\morefit\ uses the \texttt{Xoshiro128++} generator for this approach which has an initial state of 16 bytes.
Every work item (or thread for execution on the CPU) needs its own pseudo random number generator with its own internal state, 
which is centrally seeded from the host using the generators' jump function. 
For the generation on GPU this process can become prohibitively computationally expensive when a separate generator is used for each event.
Therefore, events are generated in groups, and the maximum amount of work items is set to a configurable limit which fully exploits the possible parallelism. 
Moving the generation fully to the accelerator generally improves performance as it minimises data transfers between host and accelerator. 
It is therefore the default approach in \morefit. 

\section{Usage and benchmarking}
\subsection{Benchmarking setup}
\label{sec:benchmarking}
To study the speed of the \morefit\ fitting framework (\texttt{v0.1}) it is benchmarked against RooFit \texttt{v6.32.08} and zfit \texttt{v0.24.2}. 
As benchmarking system an AMD 7950X3D 16 Core with 64GB of main memory is used.
The GPU used in the benchmarks is a NVIDIA Titan V 12GB, which allows for General Purpose GPU calculations via both OpenCL and NVIDIAs CUDA.
While the GPU is an older model based on the Volta architecture,
it has good performance for calculation at double precision (theoretical maximum of $7.45$\,TFlops),
which is crucial for the unbinned maximum likelihood fits presented here.\footnote{For applications requiring double precision the FP64 performance should be carefully checked before purchasing decisions. Compared to single precision typical reductions in double performance can correspond to a factor of 64 for consumer cards.} 

Benchmarking of course depends on the specific type of problem, two illustrative examples are studied below:
First, a simple one-dimensional mass fit of a model consisting of the sum of a Gaussian and an exponential (corresponding to Eq.~\ref{eq:massmodel}) is performed in Sec.~\ref{sec:massfit}. 
Secondly, an angular fit using Eq.~\ref{eq:angulardecayrate} is studied in Sec.~\ref{sec:angularfit}. 
To benchmark the speed of the algorithms pseudoexperiments are performed for different statistics, 
corresponding to $N=\{1000,10\,000,100\,000,1\,000\,000\}$ events. 
Each pseudoexperiment consists of the generation and subsequent fit of pseudodata, followed by the determination of the uncertainty via the \texttt{HESSE} algorithm.
The minimisation uses as starting point the generated values. 
The time for the uncertainty determination via \texttt{HESSE} is included in the total time.
For each $N$, 100 samples of pseudodata are generated and fit and the total time needed is recorded, from which the time per pseudoexperiment is calculated. 
This is repeated 10 times for each $N$ and the mean time per pseudoexperiment is reported. 
All fitting algorithms use \texttt{Minuit2} with the same Minuit strategy. 
The full code used for the benchmarking is available online in the \morefit\ repository~\cite{morefitrepo}. 

\subsection{Example I: Mass fit (nonlinear fit with 4 parameters)}
\label{sec:massfit}
\begin{figure}
  \centering
\includegraphics[width=0.7\linewidth]{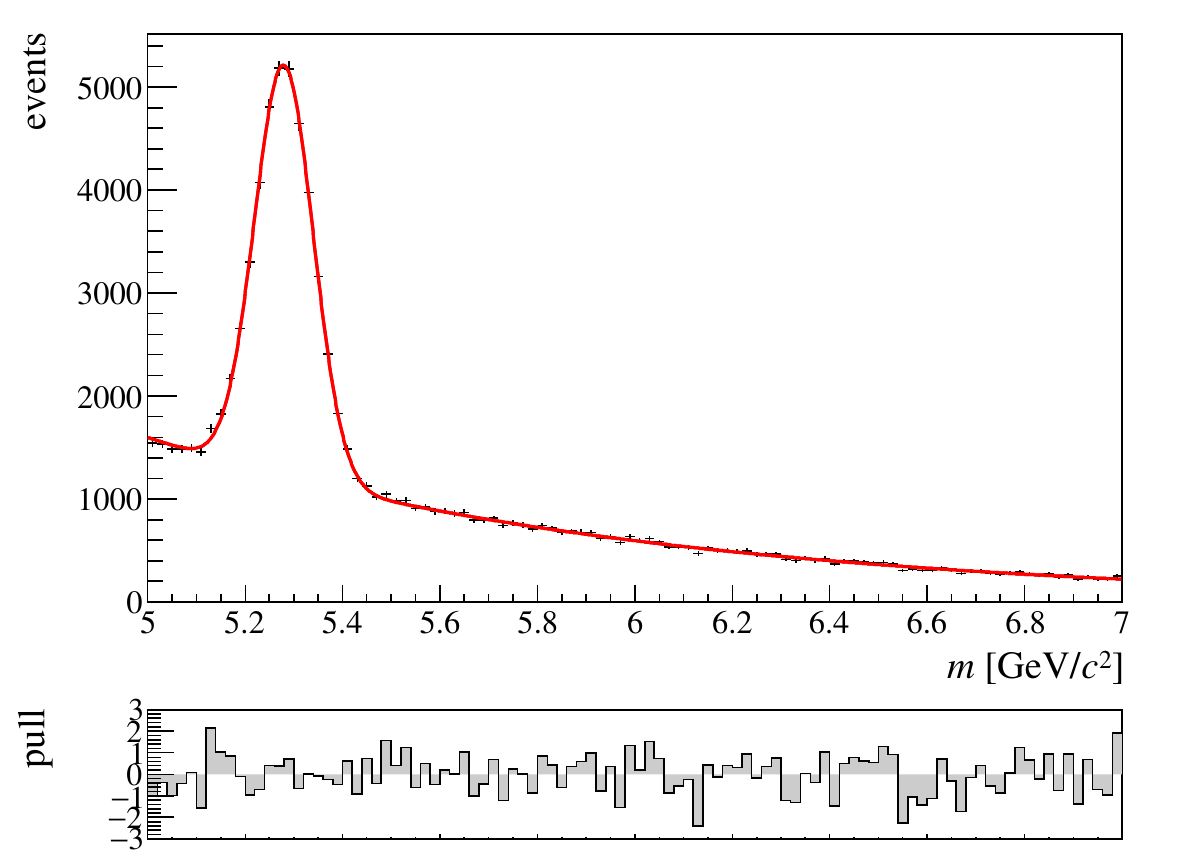}
  \caption{Distribution for pseudodata corresponding to 100\,000 events, overlaid with the fitted PDF.\label{fig:massexample}}
\end{figure}

To study the performance of the fitting frameworks on a simple example with likely highly optimised PDFs the first scenario is
the fit of 
the one-dimensional model given by Eq.~\ref{eq:massmodel}. 
In total the fit determines the four parameters $\bm{\lambda}=\{\fsig,\mu,\sigma_m,\alpha\}$, where $\fsig$ is the signal fraction, $\mu$ the mean of the Gaussian, $\sigma_m$ its width, and $\alpha$ the slope of the exponential.
In addition, the covariance matrix for the parameters is determined,
as typical for the validation of the coverage. 
The parameters used in the generation are given in Tab.~\ref{tab:massparameters}. 
The generation of pseudodata and the subsequent fit is performed in the mass range $[5,7]\gevcc$.
The pseudodata for a single experiment, overlaid with the fitted PDF is shown in Fig.~\ref{fig:massexample}. 
\begin{table}\centering
  \begin{tabular}{lr}\hline
    Parameter & Value \\ \hline\hline
    \fsig & $0.3$\\
    $\mu$ & $5.28\gevcc$ \\
    $\sigma_m$ & $0.06\gevcc$\\
    $\alpha$ & $-1.0\,{\rm GeV}^{-1}c^2$ \\
  \hline\end{tabular}
  \caption{Parameter values used in the generation for the mass fit.\label{tab:massparameters}}
\end{table}

The plotting facilities provided by \morefit\ also exploit parallelism.
In this case the computationally expensive task is the integration of the PDF to be plotted, typically with a much finer binning than the histogram over which it is to be overlaid.
For Fig.~\ref{fig:massexample} the PDF is integrated using a binning ten times finer than the histogram of the pseudodata,
this factor can be configured by the user. 
The integrals are performed on the accelerator,
parallelising over the PDF bins and 
using the analytic formulae provided by the PDFs. 

The \morefit\ usage is very similar to other \texttt{C++}-based frameworks.
Models are constructed out of PDF components, 
in this case a \texttt{SumPDF} combining a \texttt{GaussianPDF} and an \texttt{ExponentialPDF}. 
The code listing for an example \morefit\ control file is given in Listing~\ref{list:masscontrolfile} in the Appendix. 
Each PDF is derived from an abstract base class and implements computation graphs for the unnormalised probability distribution and the analytic integrals for its normalisation.
In addition, the maximum probability is provided for generation via the accept/reject method.
\morefit\ writes the required kernels from the supplied PDFs automatically, 
incorporating the requested automatic optimisations 
without requiring manual intervention by the user. 
For benchmarking purposes identical PDFs are implemented using the RooFit and zfit fitting frameworks.

\begin{figure}
\centering
  \subfloat[Numerical derivatives\label{fig:massnumerical}]{\includegraphics[width=0.775\linewidth]{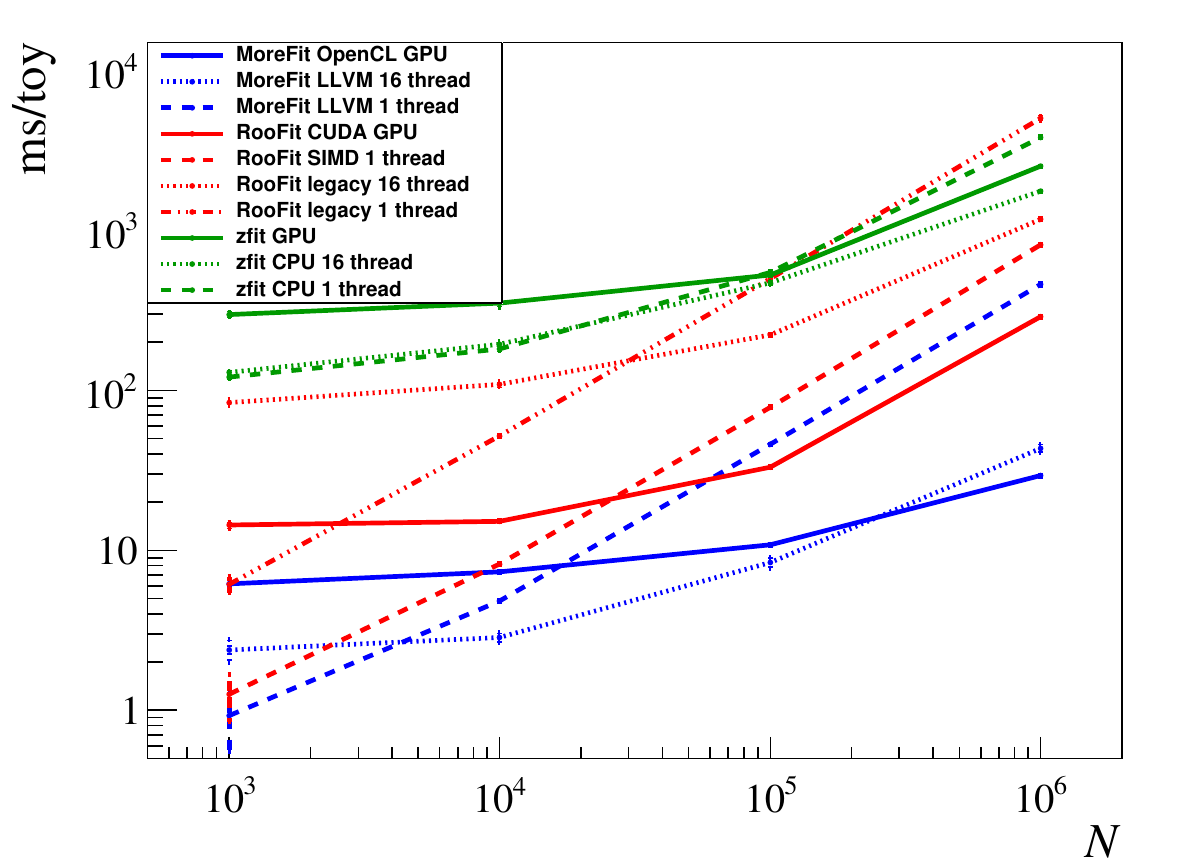}}\\
  \subfloat[Comparison with analytic derivatives\label{fig:massanalytic}]{\includegraphics[width=0.775\linewidth]{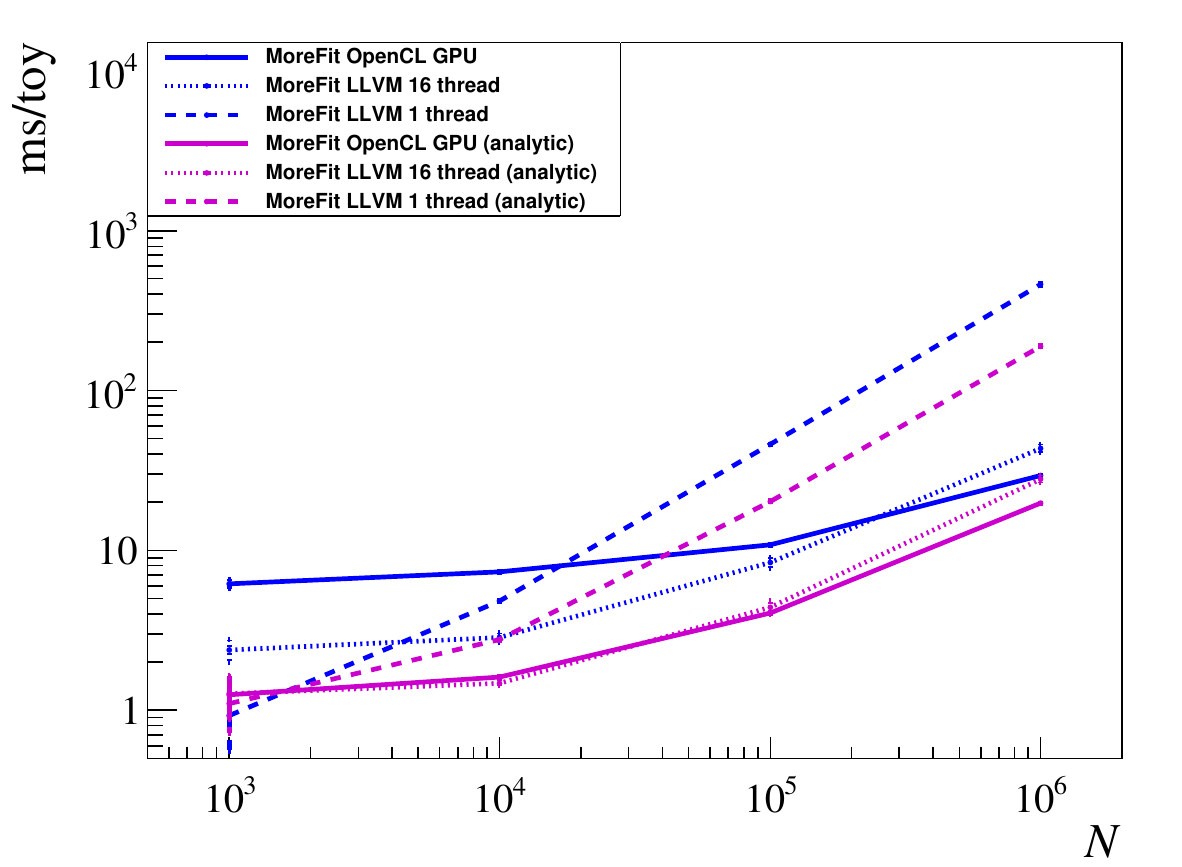}}
\caption{Time in $\mathrm{ms}$ per pseudoexperiment for a simple mass fit for different numbers of events ranging from $10^3$ to $10^6$. 
  Different colours indicate the different fitting frameworks RooFit (red), zfit (green) and \morefit~(blue). 
  The different frameworks are compared in (a) using numerical derivatives. 
  In (b)~the performance between \morefit\ using numerical (blue) and analytic (purple) gradient and Hessian matrix is compared. 
  Solid lines denote the use of the GPU, dashed (dotted) lines the use of the CPU with 1 thread (16 threads).
  \label{fig:massbench}}
\end{figure}

The benchmarking results for the mass fit study are shown in Fig.~\ref{fig:massbench} and numerical results are given in Tab.~\ref{tab:massbench}. 
A detailed breakdown of the time spent by \morefit\ in the generation, minimisation, and the determination of the uncertainties using the Hesse algorithm is given in Tab.~\ref{tab:massfit_breakdown} in the Appendix. 
In Fig.~\ref{fig:massnumerical} different colours indicate the different fitting frameworks Roofit (red), zfit (green) and \morefit~(blue). 
Solid lines denote the use of the GPU, 
dashed (dotted) lines the use of the CPU with 1 thread (16 threads). 
The RooFit SIMD backend is denoted using long dashed lines. 
Figure~\ref{fig:massanalytic} compares the benchmarking results for \morefit\ when using (blue) the numerical and (purple) the analytic gradient and Hessian matrix. 
With the numerical derivatives \morefit\ needs on average around 85 function calls to converge,
with the analytic derivatives it converges in only $2\text{--}3$ iterations, where each iteration includes evaluation of the likelihood, the gradient and the Hessian. 
Recently, RooFit added the option of a codegen backend that allows the determination of the gradient through automatic differentiation~(AD) using clad~\cite{Vassilev:2015rba,Singh:2023,Singh:2024}. 
In this benchmark using the codegen backend did not lead to an improvement in performance, therefore results for RooFit using AD are not reported. 

Comparing first the performance on the CPU with 1 thread and numerical derivatives \morefit\ performs well and is faster than the RooFit SIMD implementation 
by a factor of around $1.8$ for $N\geq 10\,000$. 
For $N=1\,000$ the compilation time of the kernel becomes significant and \morefit\ is only faster by a factor of around $1.4$.
With the analytic derivatives \morefit\ further gains a factor of around $2.4$ in speed at high statistics,
at low statistics the gains are smaller and for $N=1\,000$ 
the numerical derivatives are faster. 
This is caused by the overhead of multiple more complex kernels that need to be compiled when using analytic derivatives. 
\morefit\ scales well with the number of threads;
At high statistics the gain from using 16 threads corresponds to an increase in speed by up to an order of magnitude,
however, at the lowest statistics the overhead of starting multiple threads results in slightly worse performance. 
Unfortunately, the RooFit SIMD implementation does not yet allow for multithreading so no numbers can be reported for multithreading using this backend. 
The legacy RooFit implementation and zfit are not competitive in this benchmark. 
When comparing the performance on the GPU \morefit\ also performs very well,
and is faster by nearly an order of magnitude compared to RooFit's CUDA implementation at the highest statistics. 
On the GPU \morefit\ generally gains by using the analytic derivatives, even at low statistics.
This is due to the fact that when using analytic derivatives far fewer parameter steps need to be evaluated since the fit convergence is faster.
There is therefore much lower overhead from the kernel submission. 
The CUDA backend of zfit is significantly slower than RooFit's CUDA implementation in this benchmark. 

\begin{table}
\centering
\begin{tabular}{lR{1.5cm}R{1.5cm}R{1.5cm}R{1.5cm}}\hline
\multicolumn{5}{c}{Time [ms] per toy}\\
 & 1k & 10k & 100k & 1M \\ \hline\hline
MoreFit OpenCL GPU & $6.18$ & $7.32$ & $10.8$ & $29.3$ \\
MoreFit LLVM 16 thread & $2.37$ & $2.85$ & $8.38$ & $43.5$ \\
MoreFit LLVM 1 thread & $0.924$ & $4.80$ & $46.0$ & $461$ \\\hline
MoreFit OpenCL GPU (analytic) & $1.24$ & $1.62$ & $4.06$ & $19.7$ \\
MoreFit LLVM 16 thread (analytic) & $1.27$ & $1.47$ & $4.41$ & $28.0$ \\
MoreFit LLVM 1 thread (analytic) & $1.10$ & $2.76$ & $20.2$ & $189$ \\\hline
RooFit CUDA GPU & $14.4$ & $15.2$ & $33.2$ & $289$ \\
RooFit SIMD 1 thread & $1.26$ & $8.22$ & $78.4$ & $814$ \\
RooFit legacy 16 thread & $84.1$ & $109$ & $222$ & $1\,177$ \\
RooFit legacy 1 thread & $6.14$ & $52.1$ & $501$ & $5\,060$ \\\hline
zfit CUDA GPU & $299$ & $351$ & $530$ & $2\,529$ \\
zfit 16 thread & $129$ & $194$ & $472$ & $1\,764$ \\
zfit 1 thread & $121$ & $182$ & $548$ & $3\,835$ \\
\hline\end{tabular}
\caption{Time in $\mathrm{ms}$ per pseudoexperiment for a simple mass fit for different numbers of events ranging from $10^3$ to $10^6$.
For \morefit, the times are given for both the numerical and, denoted as analytic, the analytic gradient and Hessian.
  \label{tab:massbench}}
\end{table}

\subsection{Example II: Angular fit (linear fit with 8 parameters)}
\label{sec:angularfit}The second benchmarking scenario is a multidimensional angular fit using the differential angular decay rate given by Eq.~\ref{eq:angulardecayrate}. 
This example is chosen to compare the performance of the frameworks on a PDF which is not built-in but instead user-supplied.
The PDF is implemented in all three frameworks, Listing~\ref{list:userpdf} in the Appendix gives the \morefit\ implementation.
The RooFit implementation uses the \texttt{RooClassFactory} facilities as a base,
the zfit implementation follows the example on custom models in the zfit documentation. 
All implementations are provided with the analytic integral required for normalisation.

The fit determines the parameters $\bm{\lambda}=\{F_{\rm L}, S_3, S_4, S_5, A_{\rm FB}, S_7, S_8, S_9\}$.
In the generation of pseudoexperiments, the parameter $F_{\rm L}$ is set to $F_{\rm L}=0.6$, all other angular observables are set to zero. 
In this example, \morefit\ uses the automatic optimisation of event-dependent terms discussed in Sec.~\ref{sec:pereventoptimisation} throughout. 
This optimisation is found to improve performance in this example by up to a factor $2.5$.

\begin{figure}
\centering
\subfloat[Numerical derivatives\label{fig:kstarmumunumerical}]{\includegraphics[width=0.775\linewidth]{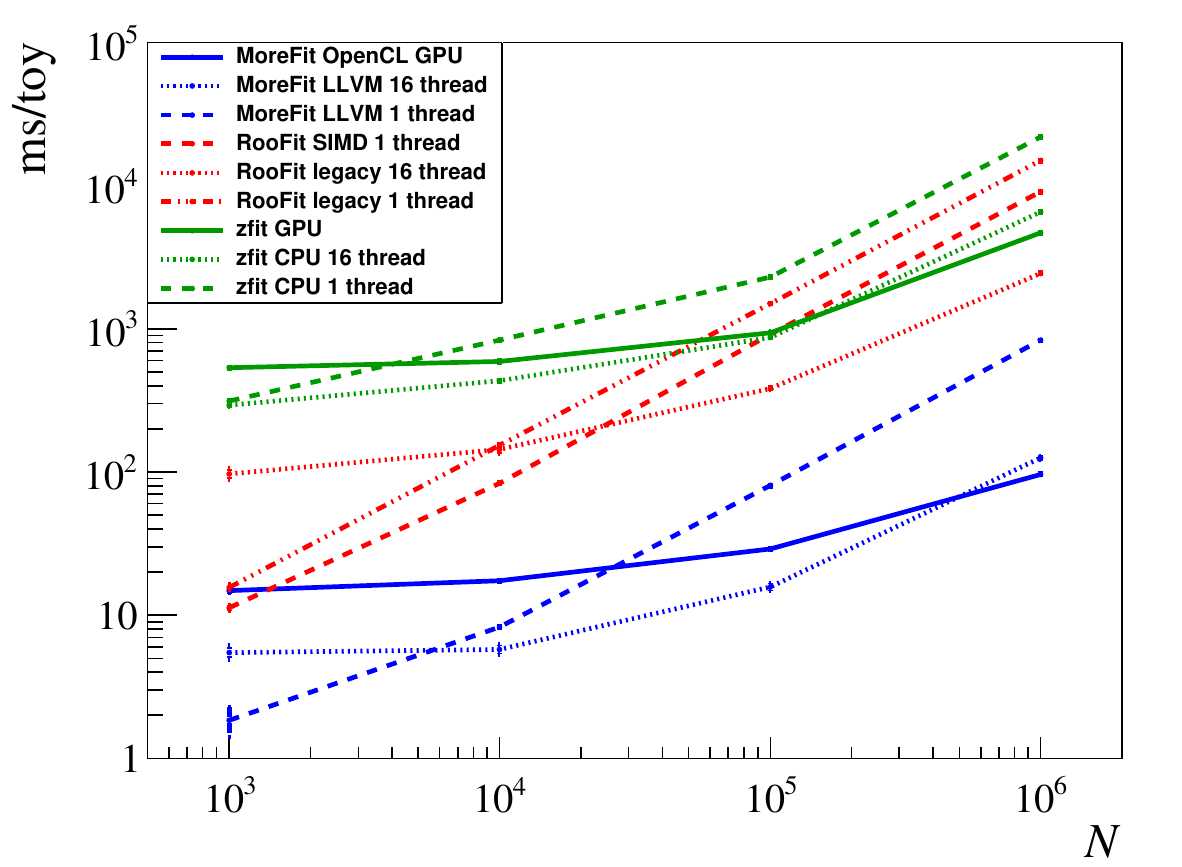}}\\
\subfloat[Comparison with analytic derivatives\label{fig:kstarmumuanalytic}]{\includegraphics[width=0.775\linewidth]{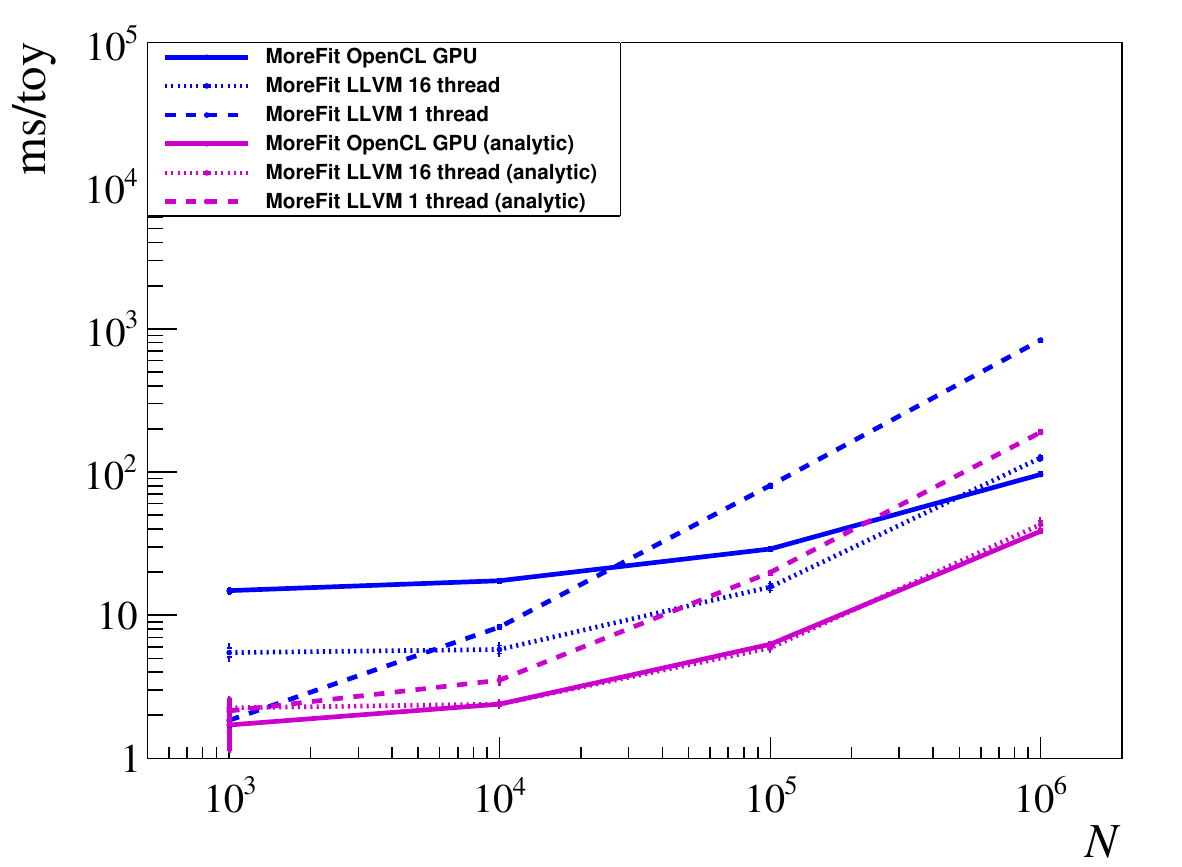}}
\caption{
  Time in $\mathrm{ms}$ per pseudoexperiment for a simple angular fit of $B^0\to K^{*0}\mu^+\mu^-$ events for different numbers of events ranging from $10^3$ to $10^6$. 
  Different colours indicate the different fitting frameworks RooFit (red), zfit (green) and \morefit~(blue).
  The different frameworks are compared in (a) using numerical derivatives.
  In (b)~the performance between \morefit\ using numerical (blue) and analytic (purple) gradient and Hessian matrix is compared. 
  Solid lines denote the use of the GPU, dashed (dotted) lines the use of the CPU with 1 thread (16 threads). 
  \label{fig:kstarbench}}
\end{figure}

The benchmarking results for the angular fit are shown in Fig.~\ref{fig:kstarbench} and numerical results are given in Tab.~\ref{tab:kstarbench}. 
A detailed breakdown of the time spent by \morefit\ in the generation, minimisation, and the determination of the uncertainties using the Hesse algorithm is given in Tab.~\ref{tab:kstarmumufit_breakdown} in the Appendix. 
The different colours in Fig.~\ref{fig:kstarmumunumerical} again indicate the different fitting frameworks Roofit (red), zfit (green) and \morefit~(blue). 
Solid lines denote the use of the GPU, 
dashed (dotted) lines the use of the CPU with 1 thread (16 threads). 
In Fig.~\ref{fig:kstarmumuanalytic} the \morefit\ results use (blue) the numerical and (purple) the analytic gradient and Hessian matrix.
With the numerical derivatives \morefit\ needs on average around 200 function calls to converge,
with the analytic derivatives it again converges in only $2\text{--}3$ iterations. 
Comparing single-threaded performance on the CPU, \morefit\ is a factor $6.6$ faster than the RooFit SIMD backend at low statistics and around a factor $11$ faster at intermediate and high statistics. 
The RooFit legacy backend and the zfit performance on the CPU is not competitive.
The acceleration of the angular PDF with the RooFit CUDA backend requires including it in the batchcompute library and thus patching the roofit sources.
This is beyond the scope of this paper and therefore no results are reported for the RooFit CUDA backend in this benchmark. 
On the GPU, \morefit\ is faster than zfit by a factor $32$\text{--}$48$ in this benchmark, 
depending on statistics.
When \morefit\ is using analytic derivatives its performance in this benchmark further improves.
On the CPU, the performance increases by up to a factor $4.5$ at high statistics and only reduces slightly at the lowest statistics when using single-threading. 
On the GPU, the performance increases by at least a factor $2.5$, with relatively faster execution at low statistics.
This is because the overhead from kernel submission can be significant for the OpenCL backend, 
and fewer parameter points need to be iterated over when analytic derivatives are supplied. 
At high statistics, the relative cost of kernel submission compared to the kernel run time reduces,
resulting in a smaller, though still very significant, advantage for the analytic derivatives. 

\begin{table}
\centering
\begin{tabular}{lR{1.5cm}R{1.5cm}R{1.5cm}R{1.5cm}}\hline
\multicolumn{5}{c}{Time [ms] per toy}\\
 & 1k & 10k & 100k & 1M \\ \hline\hline
MoreFit OpenCL GPU & $14.8$ & $17.3$ & $29.0$ & $96.5$ \\
MoreFit LLVM 16 thread & $5.48$ & $5.76$ & $15.7$ & $125$ \\
MoreFit LLVM 1 thread & $1.84$ & $8.27$ & $80.4$ & $832$ \\\hline
MoreFit OpenCL GPU (analytic) & $1.70$ & $2.39$ & $6.25$ & $38.7$ \\
MoreFit LLVM 16 thread (analytic) & $2.25$ & $2.40$ & $5.90$ & $42.9$ \\
MoreFit LLVM 1 thread (analytic) & $2.14$ & $3.51$ & $19.9$ & $190$ \\\hline
RooFit SIMD 1 thread & $11.3$ & $83.6$ & $913$ & $9\,073$ \\
RooFit legacy 16 thread & $97.4$ & $144$ & $385$ & $2\,449$ \\
RooFit legacy 1 thread & $15.7$ & $154$ & $1\,506$ & $14\,986$ \\\hline
zfit CUDA GPU & $535$ & $593$ & $943$ & $4\,694$ \\
zfit 16 thread & $293$ & $435$ & $873$ & $6\,571$ \\
zfit 1 thread & $313$ & $832$ & $2\,305$ & $21\,940$ \\
\hline\end{tabular}
\caption{Time in $\mathrm{ms}$ per pseudoexperiment for a simple angular fit of $B^0\to K^{*0}\mu^+\mu^-$ events for different numbers of events ranging from $10^3$ to $10^6$.
For \morefit, the times are given for both the numerical and, denoted as analytic, the analytic gradient and Hessian.
  \label{tab:kstarbench}}
\end{table}

\section{Summary and outlook}
\label{sec:summary}
This article introduces \morefit, a new fitting framework focused on maximally exploiting parallelism on heterogeneous platforms. 
\morefit\ is still at an early stage in its development. Nevertheless, it already shows impressive performance.
Some of the benchmarks discussed above show increases in speed of an order of magnitude or more. 
This demonstrates that the \morefit\ approach of using computation graphs that are automatically optimised for the different problems in parameter fits can indeed be quite powerful. 
The fact that the automatic optimisations discussed here can be applied without requiring manual interventions make this a very attractive approach also for user-supplied PDFs. 
The discussed optimisation techniques 
can enable or make more robust computationally otherwise prohibitively expensive statistical techniques, for example in the area of coverage correction. 
By relying on open and vendor-independent standards \morefit\ aims furthermore to be as broadly useable as possible. 
As \morefit\ is still at an early stage in its development there are several features missing that are obvious areas for future improvement.
The library of built-in PDFs is still quite small and will be extended in future releases.
Furthermore, efficient general purpose acceptance corrections will be an essential future addition to the \morefit\ feature set.
Finally, binned fits will be explored, which are currently not supported.
In summary, \morefit\ presents a new unique approach to the determination of parameters in unbinned maximum likelihood fits which is showing very promising performance. 
It demonstrates that there are still significant improvements possible in this area with respect to 
speed and efficiency, allowing for a more sustainable use of limited computational resources. 

\section*{Acknowledgements}
C.\,L.\ gratefully acknowledges support by the Heisenberg programme of the Deutsche Forschungsgemeinschaft (DFG), grant identifier LA 3937/2-1.

\clearpage

\setboolean{inbibliography}{true}
\bibliographystyle{JHEP}
\bibliography{main}

\providecommand{\href}[2]{#2}\begingroup\raggedright\begin{thebibliography}{10}

\bibitem{James:2006zz}
F.~James, \emph{{Statistical methods in experimental physics}}. Hackensack,
  USA: World Scientific (2006) 345 p, 2006.

\bibitem{Neyman:1937uhy}
J.~Neyman, \emph{{Outline of a Theory of Statistical Estimation Based on the
  Classical Theory of Probability}},
  \href{https://doi.org/10.1098/rsta.1937.0005}{\emph{Phil. Trans. Roy. Soc.
  Lond. A} {\bfseries 236} (1937) 333}.

\bibitem{Feldman:1997qc}
G.~J. Feldman and R.~D. Cousins, \emph{{A Unified approach to the classical
  statistical analysis of small signals}},
  \href{https://doi.org/10.1103/PhysRevD.57.3873}{\emph{Phys. Rev. D}
  {\bfseries 57} (1998) 3873}
  [\href{https://arxiv.org/abs/physics/9711021}{{\ttfamily physics/9711021}}].

\bibitem{Verkerke:2003ir}
W.~Verkerke and D.~P. Kirkby, \emph{{The RooFit toolkit for data modeling}},
  {\emph{eConf} {\bfseries C0303241} (2003) MOLT007}
  [\href{https://arxiv.org/abs/physics/0306116}{{\ttfamily physics/0306116}}].

\bibitem{Eschle:2019jmu}
J.~Eschle, A.~Puig~Navarro, R.~Silva~Coutinho and N.~Serra, \emph{{zfit:
  scalable pythonic fitting}},
  \href{https://arxiv.org/abs/1910.13429}{{\ttfamily 1910.13429}}.

\bibitem{Andreassen_2014}
R.~Andreassen, B.~T. Meadows, M.~de~Silva, M.~D. Sokoloff and K.~Tomko,
  \emph{Goofit: A library for massively parallelising maximum-likelihood fits},
  \href{https://doi.org/10.1088/1742-6596/513/5/052003}{\emph{Journal of
  Physics: Conference Series} {\bfseries 513} (2014) 052003}.

\bibitem{Schreiner:2017csm}
H.~Schreiner, C.~Hasse, B.~Hittle, H.~Pandey, M.~Sokoloff and K.~Tomko,
  \emph{{GooFit 2.0}},
  \href{https://doi.org/10.1088/1742-6596/1085/4/042014}{\emph{J. Phys. Conf.
  Ser.} {\bfseries 1085} (2018) 042014}
  [\href{https://arxiv.org/abs/1710.08826}{{\ttfamily 1710.08826}}].

\bibitem{TensorFlowAnalysis}
A.~Poluektov et~al., ``{TensorFlowAnalysis}.''
  \url{https://gitlab.cern.ch/poluekt/TensorFlowAnalysis/}.

\bibitem{Nickolls:2008gqs}
J.~Nickolls, I.~Buck, M.~Garland and K.~Skadron, \emph{{Scalable Parallel
  Programming with CUDA}},
  \href{https://doi.org/10.1145/1365490.1365500}{\emph{Queue} {\bfseries 6}
  (2008) 40}.

\bibitem{tensorflow2015-whitepaper}
M.~Abadi, A.~Agarwal, P.~Barham, E.~Brevdo, Z.~Chen, C.~Citro et~al.,
  \emph{{TensorFlow}: Large-scale machine learning on heterogeneous systems},
  2015.

\bibitem{OpenCL}
J.~E. Stone, D.~Gohara and G.~Shi, \emph{Opencl: A parallel programming
  standard for heterogeneous computing systems},
  \href{https://doi.org/10.1109/MCSE.2010.69}{\emph{Computing in Science \&
  Engineering} {\bfseries 12} (2010) 66}.

\bibitem{LLVM:CGO04}
C.~Lattner and V.~Adve, \emph{{LLVM}: A compilation framework for lifelong
  program analysis and transformation},  in \emph{CGO}, (San Jose, CA, USA),
  pp.~75--88, Mar, 2004.

\bibitem{clang}
{The Clang community}, ``{Clang: a C language family frontend for LLVM}.''
  \url{https://clang.llvm.org/}, 2025.

\bibitem{Brun:1997pa}
R.~Brun and F.~Rademakers, \emph{{ROOT: An object oriented data analysis
  framework}}, \href{https://doi.org/10.1016/S0168-9002(97)00048-X}{\emph{Nucl.
  Instrum. Meth.} {\bfseries A389} (1997) 81}.

\bibitem{James:1975dr}
F.~James and M.~Roos, \emph{{Minuit: A System for Function Minimization and
  Analysis of the Parameter Errors and Correlations}},
  \href{https://doi.org/10.1016/0010-4655(75)90039-9}{\emph{Comput. Phys.
  Commun.} {\bfseries 10} (1975) 343}.

\bibitem{James:1994vla}
F.~James, \emph{{MINUIT Function Minimization and Error Analysis: Reference
  Manual Version 94.1}},  Tech. Rep. CERN-D-506, CERN-D506, CERN, 1994.

\bibitem{FletcherPowell}
R.~Fletcher and M.~J.~D. Powell, \emph{A rapidly convergent descent method for
  minimization}, \href{https://doi.org/10.1093/comjnl/6.2.163}{\emph{The
  Computer Journal} {\bfseries 6} (1963) 163}
  [\href{https://arxiv.org/abs/https://academic.oup.com/comjnl/article-pdf/6/2/163/1041527/6-2-163.pdf}{{\ttfamily
  https://academic.oup.com/comjnl/article-pdf/6/2/163/1041527/6-2-163.pdf}}].

\bibitem{Broyden}
C.~G. Broyden, \emph{{The Convergence of a Class of Double-rank Minimization
  Algorithms 1. General Considerations}},
  \href{https://doi.org/10.1093/imamat/6.1.76}{\emph{Ima J. Appl. Math.}
  {\bfseries 6} (1970) 76}.

\bibitem{Fletcher}
R.~Fletcher, \emph{A new approach to variable metric algorithms},
  \href{https://doi.org/10.1093/comjnl/13.3.317}{\emph{The Computer Journal}
  {\bfseries 13} (1970) 317}
  [\href{https://arxiv.org/abs/https://academic.oup.com/comjnl/article-pdf/13/3/317/988678/130317.pdf}{{\ttfamily
  https://academic.oup.com/comjnl/article-pdf/13/3/317/988678/130317.pdf}}].

\bibitem{Goldfarb}
D.~Goldfarb, \emph{A family of variable-metric methods derived by variational
  means}, {\emph{Mathematics of Computation} {\bfseries 24} (1970) 23}.

\bibitem{Shanno}
D.~F. Shanno, \emph{Conditioning of quasi-newton methods for function
  minimization}, {\emph{Mathematics of Computation} {\bfseries 24} (1970) 647}.

\bibitem{Langenbruch:2019nwe}
C.~Langenbruch, \emph{{Parameter uncertainties in weighted unbinned maximum
  likelihood fits}},
  \href{https://doi.org/10.1140/epjc/s10052-022-10254-8}{\emph{Eur. Phys. J. C}
  {\bfseries 82} (2022) 393}
  [\href{https://arxiv.org/abs/1911.01303}{{\ttfamily 1911.01303}}].

\bibitem{KahanSummation}
W.~Kahan, \emph{Pracniques: further remarks on reducing truncation errors},
  \href{https://doi.org/10.1145/363707.363723}{\emph{Commun. ACM} {\bfseries 8}
  (1965) 40}.

\bibitem{LHCb:2015svh}
{\scshape LHCb} collaboration, \emph{{Angular analysis of the $B^{0} \to K^{*0}
  \mu^{+} \mu^{-}$ decay using 3 fb$^{-1}$ of integrated luminosity}},
  \href{https://doi.org/10.1007/JHEP02(2016)104}{\emph{JHEP} {\bfseries 02}
  (2016) 104} [\href{https://arxiv.org/abs/1512.04442}{{\ttfamily
  1512.04442}}].

\bibitem{LHCb:2020lmf}
{\scshape LHCb} collaboration, \emph{{Measurement of $CP$-Averaged Observables
  in the $B^{0}\rightarrow K^{*0}\mu^{+}\mu^{-}$ Decay}},
  \href{https://doi.org/10.1103/PhysRevLett.125.011802}{\emph{Phys. Rev. Lett.}
  {\bfseries 125} (2020) 011802}
  [\href{https://arxiv.org/abs/2003.04831}{{\ttfamily 2003.04831}}].

\bibitem{Altmannshofer:2008dz}
W.~Altmannshofer, P.~Ball, A.~Bharucha, A.~J. Buras, D.~M. Straub and M.~Wick,
  \emph{{Symmetries and Asymmetries of $B \to K^{*} \mu^{+} \mu^{-}$ Decays in
  the Standard Model and Beyond}},
  \href{https://doi.org/10.1088/1126-6708/2009/01/019}{\emph{JHEP} {\bfseries
  01} (2009) 019} [\href{https://arxiv.org/abs/0811.1214}{{\ttfamily
  0811.1214}}].

\bibitem{MersenneTwister}
M.~Matsumoto and T.~Nishimura, \emph{Mersenne twister: a 623-dimensionally
  equidistributed uniform pseudo-random number generator},
  \href{https://doi.org/10.1145/272991.272995}{\emph{ACM Trans. Model. Comput.
  Simul.} {\bfseries 8} (1998) 3–30}.

\bibitem{Xoshiro}
D.~{Blackman} and S.~{Vigna}, \emph{{Scrambled Linear Pseudorandom Number
  Generators}}, \href{https://doi.org/10.48550/arXiv.1805.01407}{\emph{arXiv
  e-prints} (2018) arXiv:1805.01407}
  [\href{https://arxiv.org/abs/1805.01407}{{\ttfamily 1805.01407}}].

\bibitem{oneill:pcg2014}
M.~E. O'Neill, \emph{{PCG: A Family of Simple Fast Space-Efficient
  Statistically Good Algorithms for Random Number Generation}},  Tech. Rep.
  HMC-CS-2014-0905, Harvey Mudd College, Claremont, CA, Sept., 2014.

\bibitem{morefitrepo}
C.~Langenbruch, ``{MoreFit v0.1}.''
  \url{https://www.github.com/langenbruch/morefit},
  \href{https://doi.org/10.5281/zenodo.17792534}{doi.org/10.5281/zenodo.17792534}.

\bibitem{Vassilev:2015rba}
V.~Vassilev, M.~Vassilev, A.~Penev, L.~Moneta and V.~Ilieva, \emph{{Clad
  \textemdash{} Automatic Differentiation Using Clang and LLVM}},
  \href{https://doi.org/10.1088/1742-6596/608/1/012055}{\emph{J. Phys. Conf.
  Ser.} {\bfseries 608} (2015) 012055}.

\bibitem{Singh:2023}
G.~Singh, J.~Rembser, L.~Moneta, D.~Lange and V.~Vassilev, \emph{Automatic
  differentiation of binned likelihoods with roofit and clad},
  \href{https://arxiv.org/abs/2304.02650}{{\ttfamily 2304.02650}}.

\bibitem{Singh:2024}
{Singh, Garima}, {Rembser, Jonas}, {Moneta, Lorenzo}, {Lange, David} and
  {Vassilev, Vassil}, \emph{Making likelihood calculations fast: Automatic
  differentiation applied to roofit},
  \href{https://doi.org/10.1051/epjconf/202429506014}{\emph{EPJ Web of Conf.}
  {\bfseries 295} (2024) 06014}.

\end{thebibliography}\endgroup

\clearpage

\appendix

\section{Appendix}
\label{sec:appendix}

\begin{figure}[h]
  \centering
  \includegraphics[width=0.95\linewidth]{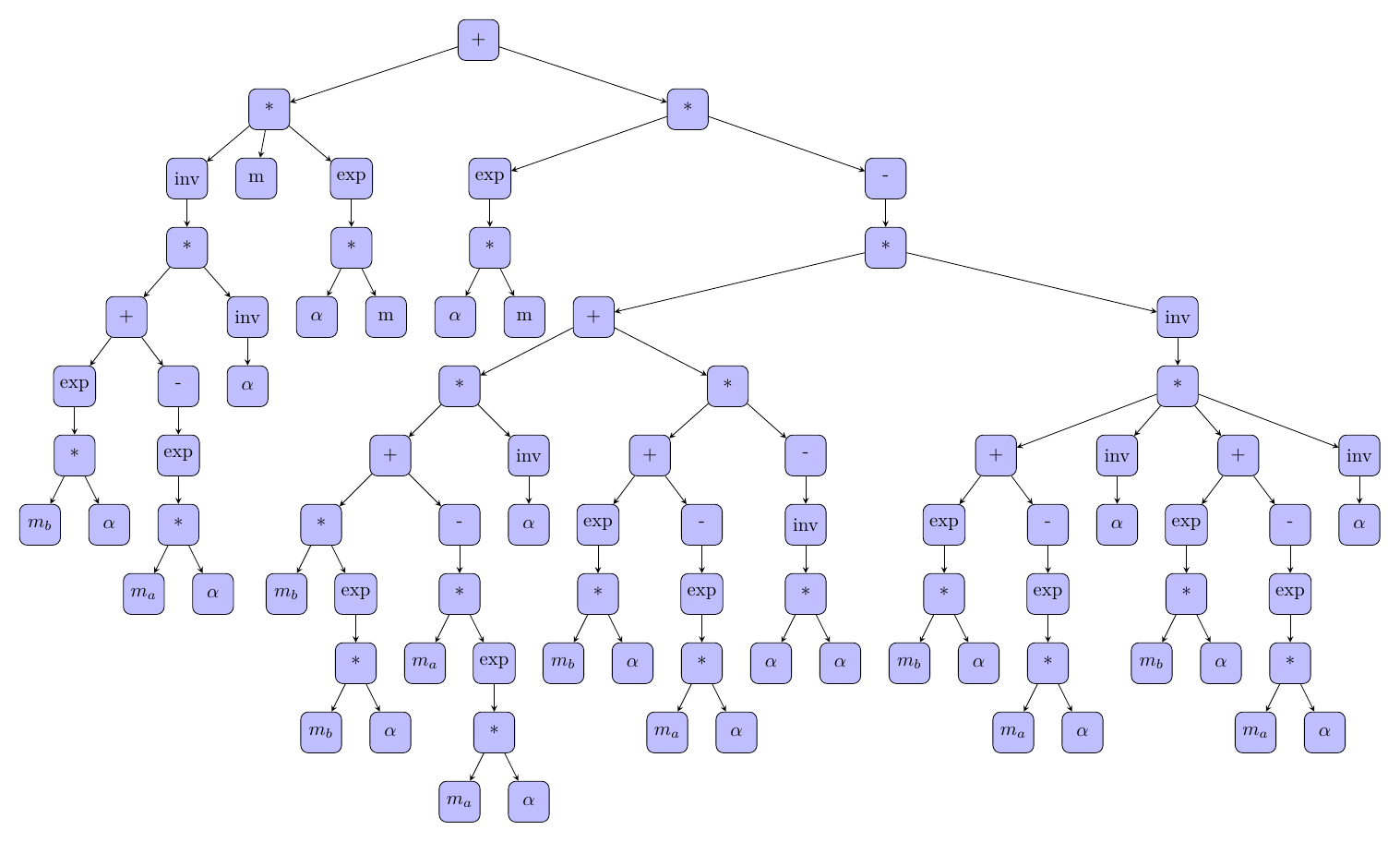}
  \caption{Computation graph for the derivative of a simple normalised exponential PDF with respect to the slope parameter $\alpha$. Note that the graph is slightly simplified for illustration purposes, as \morefit\ by default in addition handles the special case $\alpha=0$.\label{fig:derivativeexponential}}
\end{figure}

\begin{table}[h]
  \centering
  \begin{tabular}{lrrr}\hline
    & \multicolumn{3}{c}{Fraction of time [\%]}\\
    & \parbox{\widthof{Hesse}}{\hfill Gen.} & \parbox{\widthof{Hesse}}{\hfill Min.} & \parbox{\widthof{Hesse}}{Hesse}\\ \hline\hline
    MoreFit LLVM 1 thread & 12 & 70 & 18 \\
    MoreFit LLVM 16 thread & 14 & 68 & 17 \\
    MoreFit OpenCL & 32 & 56 & 12 \\ \hline
    MoreFit LLVM 1 thread (analytic) & 29 & 60 & 12 \\
    MoreFit LLVM 16 thread (analytic) & 27 & 62 & 11 \\
    MoreFit OpenCL (analytic) & 50 & 46 & 3 \\
    \hline\end{tabular}
    \caption{\small Detailed breakdown of the time spent by \morefit\ in the mass fit in Sec.~\ref{sec:massfit} on the generation, the minimisation and the post-fit run of the Hesse algorithm for pseudoexperiments containing $1\,{\rm M}$ events.\label{tab:massfit_breakdown}}
\end{table}

\begin{table}[h]
  \centering
\begin{tabular}{lrrr}\hline
& \multicolumn{3}{c}{Fraction of time [\%]}\\
& \parbox{\widthof{Hesse}}{\hfill Gen.} & \parbox{\widthof{Hesse}}{\hfill Min.} & \parbox{\widthof{Hesse}}{Hesse}\\ \hline\hline
MoreFit LLVM 1 thread & 5 & 73 & 22 \\
MoreFit LLVM 16 thread & 5 & 75 & 20 \\
MoreFit OpenCL & 15 & 67 & 18 \\ \hline
MoreFit LLVM 1 thread (analytic) & 21 & 67 & 12 \\
MoreFit LLVM 16 thread (analytic) & 16 & 75 & 9 \\
MoreFit OpenCL (analytic) & 48 & 47 & 5 \\
\hline\end{tabular}
    \caption{\small Detailed breakdown of the time spent by \morefit\ in the angular fit in Sec.~\ref{sec:angularfit} on the generation, the minimisation and the post-fit run of the Hesse algorithm for pseudoexperiments containing $1\,{\rm M}$ events.\label{tab:kstarmumufit_breakdown}}
  \end{table}

\clearpage

\begin{lstlisting}[float,caption={OpenCL kernel performing the precomputation step for the optimisation of event-dependent terms as discussed in Sec.~\ref{sec:pereventoptimisation}.},captionpos=b,label=list:precomputationkstar]
__kernel void precompute_kernel(__const int nevents, __const int nevents_padded, __global const double* data, __global double* output)
{
int idx = get_global_id(0);
const double ctl = data[idx];
const double ctk = data[nevents_padded*1+idx];
const double phi = data[nevents_padded*2+idx];
output[idx] = ctl;
output[nevents_padded*1+idx] = ctk;
output[nevents_padded*2+idx] = phi;
output[nevents_padded*3+idx] = (8.952465548919113e-02*sin((2.000000000000000e+00*phi))*(1.000000000000000e+00+-((ctl*ctl)))*(1.000000000000000e+00+-((ctk*ctk))));
output[nevents_padded*4+idx] = (3.580986219567645e-01*sin(phi)*ctk*sqrt((1.000000000000000e+00+-((ctk*ctk))))*ctl*sqrt((1.000000000000000e+00+-((ctl*ctl)))));
output[nevents_padded*5+idx] = (1.790493109783823e-01*sin(phi)*sqrt((1.000000000000000e+00+-((ctl*ctl))))*ctk*sqrt((1.000000000000000e+00+-((ctk*ctk)))));
output[nevents_padded*6+idx] = (1.193662073189215e-01*ctl*(1.000000000000000e+00+-((ctk*ctk))));
output[nevents_padded*7+idx] = (1.790493109783823e-01*cos(phi)*sqrt((1.000000000000000e+00+-((ctl*ctl))))*ctk*sqrt((1.000000000000000e+00+-((ctk*ctk)))));
output[nevents_padded*8+idx] = (3.580986219567645e-01*cos(phi)*ctk*sqrt((1.000000000000000e+00+-((ctk*ctk))))*ctl*sqrt((1.000000000000000e+00+-((ctl*ctl)))));
output[nevents_padded*9+idx] = (8.952465548919113e-02*cos((2.000000000000000e+00*phi))*(1.000000000000000e+00+-((ctl*ctl)))*(1.000000000000000e+00+-((ctk*ctk))));
output[nevents_padded*10+idx] = (8.952465548919113e-02*(-1.000000000000000e+00+(2.000000000000000e+00*ctl*ctl))*ctk*ctk);
output[nevents_padded*11+idx] = (2.238116387229778e-02*(-1.000000000000000e+00+(2.000000000000000e+00*ctl*ctl))*(1.000000000000000e+00+-((ctk*ctk))));
output[nevents_padded*12+idx] = (6.714349161689334e-02*(1.000000000000000e+00+-((ctk*ctk))));
output[nevents_padded*13+idx] = (8.952465548919113e-02*ctk*ctk);
}
\end{lstlisting}

\begin{lstlisting}[float,caption={OpenCL kernel calculating the logarithmic probability when optimising event-dependent terms as discussed in Sec.~\ref{sec:pereventoptimisation}. 
  The kernel uses as input the output of the precompute kernel in Listing~\ref{list:precomputationkstar}.},captionpos=b,label=list:pereventkstar]
__kernel void lh_kernel(__const int nevents, __const int nevents_padded, __global const double* data, __global double* output, __global const double* parameters)
{
int idx = get_global_id(0);
const double ctl = data[idx];
const double ctk = data[nevents_padded*1+idx];
const double phi = data[nevents_padded*2+idx];
const double morefit_eventbuffer_3 = data[nevents_padded*3+idx];
const double morefit_eventbuffer_4 = data[nevents_padded*4+idx];
const double morefit_eventbuffer_5 = data[nevents_padded*5+idx];
const double morefit_eventbuffer_6 = data[nevents_padded*6+idx];
const double morefit_eventbuffer_7 = data[nevents_padded*7+idx];
const double morefit_eventbuffer_8 = data[nevents_padded*8+idx];
const double morefit_eventbuffer_9 = data[nevents_padded*9+idx];
const double morefit_eventbuffer_10 = data[nevents_padded*10+idx];
const double morefit_eventbuffer_11 = data[nevents_padded*11+idx];
const double morefit_eventbuffer_12 = data[nevents_padded*12+idx];
const double morefit_eventbuffer_13 = data[nevents_padded*13+idx];
const double Fl = parameters[0];
const double S3 = parameters[1];
const double S4 = parameters[2];
const double S5 = parameters[3];
const double Afb = parameters[4];
const double S7 = parameters[5];
const double S8 = parameters[6];
const double S9 = parameters[7];
const double morefit_parambuffer_0 = parameters[8];
output[idx] = log(((morefit_eventbuffer_3*S9)+(morefit_eventbuffer_4*S8)+(morefit_eventbuffer_5*S7)+(morefit_eventbuffer_6*Afb)+(morefit_eventbuffer_7*S5)+(morefit_eventbuffer_8*S4)+(morefit_eventbuffer_9*S3)+(morefit_eventbuffer_10*-(Fl))+(morefit_eventbuffer_11*morefit_parambuffer_0)+(morefit_eventbuffer_12*morefit_parambuffer_0)+(morefit_eventbuffer_13*Fl)));
}
\end{lstlisting}  

\begin{lstlisting}[float,caption={Example control file to run a toy study as discussed in Sec.~\ref{sec:massfit}.},captionpos=b,label=list:masscontrolfile]
  //use double precision throughout
  typedef double kernelT;
  typedef double evalT;

  //options for the compute backends
  morefit::compute_options compute_opts;
  compute_opts.opencl_platform = 0;
  compute_opts.opencl_device = 0;
  compute_opts.print_kernel = true;
  compute_opts.print();

  //use OpenCL backend
  typedef morefit::OpenCLBackend backendT;
  typedef morefit::OpenCLBlock<kernelT, evalT> blockT;
  morefit::OpenCLBackend backend(&compute_opts);

  //define event variable m, parameters, and construct PDFs
  morefit::dimension<evalT> m("m", "#it{m} [GeV/#it{c}^{2}]", 5.0, 7.0, false);
  morefit::parameter<evalT> mu("mu", "m(B^{+})", 5.28, 5.0, 6.0, 0.01, false);
  morefit::parameter<evalT> sigma("sigma", "\\sigma(B^{+})", 0.06, 0.005, 0.130, 0.001, false);
  morefit::parameter<evalT> fsig("fsig", "f_{\\mathrm{sig}}", 0.3, 0.0, 1.0, 0.01, false);
  morefit::parameter<evalT> alpha("alpha", "\\alpha_{\\mathrm{bkg}}", -1.0, -10.0, 10.0, 0.01, false);  
  morefit::GaussianPDF<kernelT, evalT> gaus(&m, &mu, &sigma);
  morefit::ExponentialPDF<kernelT, evalT> exp(&m, &alpha);
  morefit::SumPDF<kernelT, evalT> sum(&gaus, &exp, &fsig);
  std::vector<morefit::parameter<evalT>*> params({&mu, &sigma, &fsig, &alpha});

  //pseudo random number generator
  morefit::Xoshiro128pp rnd;
  rnd.setSeed(229387429ULL);

  //generator options
  morefit::generator_options gen_opts;
  gen_opts.rndtype = morefit::generator_options::randomization_type::on_accelerator;
  gen_opts.print_level = 0;
  gen_opts.print();

  //fitter options
  morefit::fitter_options fit_opts;
  fit_opts.minuit_printlevel = 0;	      
  fit_opts.minimizer = morefit::fitter_options::minimizer_type::Minuit2;
  fit_opts.optimize_dimensions = false;
  fit_opts.optimize_parameters = true;
  fit_opts.analytic_gradient = false;
  fit_opts.analytic_hessian = false;
  fit_opts.kahan_on_accelerator = true;
  fit_opts.print_level = 0;
  fit_opts.analytic_fisher = false;
  fit_opts.print();

  //run toy study with 100 experiments, each with 10000 events
  morefit::toystudy<kernelT, evalT, backendT, blockT> toy(&fit_opts, &gen_opts, &backend, &rnd);
  toy.toy(&sum, params, {&m}, 100, 10000);
\end{lstlisting}

\begin{lstlisting}[float,basicstyle=\tiny\ttfamily,caption={Implementation of the PDF used for the angular fit in Sec.~\ref{sec:angularfit}. The definite integral over the angles is omitted here since it is only needed for plotting.},captionpos=b,label=list:userpdf]
  template <typename kernelT=double, typename evalT=double> 
  class KstarmumuAngularPDF: public PDF<kernelT, evalT> {
  public:
    KstarmumuAngularPDF(dimension<evalT>* ctl, dimension<evalT>* ctk, dimension<evalT>* phi, parameter<evalT>* Fl, parameter<evalT>* S3, parameter<evalT>* S4, parameter<evalT>* S5, parameter<evalT>* Afb, parameter<evalT>* S7, parameter<evalT>* S8, parameter<evalT>* S9)
    {           
      this->dimensions_ = std::vector<dimension<evalT>*>({ctl, ctk, phi});
      this->parameters_ = std::vector<parameter<evalT>*>({Fl, S3, S4, S5, Afb, S7, S8, S9});
    }      
    virtual std::unique_ptr<ComputeGraphNode<kernelT, evalT>> prob() const override //unnormalised pdf
    {
      constexpr auto Constant_ = &Constant<kernelT, evalT>; //to avoid typing template arguments
      constexpr auto Variable_ = &Variable<kernelT, evalT>;
      typedef std::unique_ptr<ComputeGraphNode<kernelT,evalT>> Ptr;
      Ptr c = Constant_(9.0/32.0/M_PI);
      Ptr costhetal = Variable_(ctl()->get_name());
      Ptr costhetak = Variable_(ctk()->get_name());
      Ptr costhetal2 = Variable_(ctl()->get_name())*Variable_(ctl()->get_name());
      Ptr costhetak2 = Variable_(ctk()->get_name())*Variable_(ctk()->get_name());
      Ptr cos2thetal = 2.0*costhetal2->copy() - 1.0;
      Ptr cos2thetak = 2.0*costhetak2->copy() - 1.0;
      Ptr sinthetal2 = 1.0 - costhetal2->copy();
      Ptr sinthetak2 = 1.0 - costhetak2->copy();
      Ptr sinthetal = Sqrt<kernelT,evalT>(sinthetal2->copy());
      Ptr sinthetak = Sqrt<kernelT,evalT>(sinthetak2->copy());
      Ptr sin2thetal = 2.0*sinthetal->copy()*Variable_(ctl()->get_name());
      Ptr sin2thetak = 2.0*sinthetak->copy()*Variable_(ctk()->get_name());
      Ptr cosphi = Cos<kernelT,evalT>(Variable_(phi()->get_name()));
      Ptr cos2phi = Cos<kernelT,evalT>(2.0*Variable_(phi()->get_name()));
      Ptr sinphi = Sin<kernelT,evalT>(Variable_(phi()->get_name()));
      Ptr sin2phi = Sin<kernelT,evalT>(2.0*Variable_(phi()->get_name()));      
      return
	  c->copy() * sinthetak2->copy() * 3.0/4.0 * (1.0-Variable_(Fl()->get_name()))
	+ c->copy() * costhetak2->copy() * Variable_(Fl()->get_name())
	+ c->copy() * sinthetak2->copy() * cos2thetal->copy() * 1.0/4.0 * (1.0-Variable_(Fl()->get_name()))
	+ c->copy() * costhetak2->copy() * cos2thetal->copy() * (-Variable_(Fl()->get_name()))
	+ c->copy() * sinthetak2->copy() * sinthetal2->copy() * cos2phi->copy() * Variable_(S3()->get_name())
	+ c->copy() * sin2thetak->copy() * sin2thetal->copy() * cosphi->copy() * Variable_(S4()->get_name())
	+ c->copy() * sin2thetak->copy() * sinthetal->copy() * cosphi->copy() * Variable_(S5()->get_name())
	+ c->copy() * sinthetak2->copy() * costhetal->copy() * 4.0/3.0 * Variable_(Afb()->get_name())
	+ c->copy() * sin2thetak->copy() * sinthetal->copy() * sinphi->copy() * Variable_(S7()->get_name())
	+ c->copy() * sin2thetak->copy() * sin2thetal->copy() * sinphi->copy() * Variable_(S8()->get_name())
	+ c->copy() * sinthetak2->copy() * sinthetal2->copy() * sin2phi->copy() * Variable_(S9()->get_name());
    }
    virtual std::unique_ptr<ComputeGraphNode<kernelT, evalT>> norm() const override //normalisation
    { return Constant<kernelT, evalT>(1.0); }
    dimension<evalT>* ctl() const //convenient access to event variables
    { return this->dimensions_.at(0); }
    dimension<evalT>* ctk() const
    { return this->dimensions_.at(1); }
    dimension<evalT>* phi() const
    { return this->dimensions_.at(2); }
    parameter<evalT>* Fl() const //convenient access to parameters
    { return this->parameters_.at(0); } 
    parameter<evalT>* S3() const
    { return this->parameters_.at(1); }
    parameter<evalT>* S4() const
    { return this->parameters_.at(2); }
    parameter<evalT>* S5() const
    { return this->parameters_.at(3); }
    parameter<evalT>* Afb() const
    { return this->parameters_.at(4); }
    parameter<evalT>* S7() const
    { return this->parameters_.at(5); }
    parameter<evalT>* S8() const
    { return this->parameters_.at(6); }
    parameter<evalT>* S9() const
    { return this->parameters_.at(7); }
    virtual evalT get_max() const override //maximum for accept/reject
    {
      return 9.0/32.0/M_PI*(4.0 + fabs(S3()->get_value()) + fabs(S4()->get_value()) + fabs(S5()->get_value()) + 4.0/3.0*fabs(Afb()->get_value()) + fabs(S7()->get_value()) + fabs(S8()->get_value()) + fabs(S9()->get_value()));
    }
  };

\end{lstlisting}

\end{document}